# DISPERSION CHAIN OF VLASOV EQUATIONS

E.E. Perepelkin[a,b,d], B.I. Sadovnikov[a], N.G. Inozemtseva[b,c], I.I. Aleksandrov[a,b]

[a] *Faculty of Physics, Lomonosov Moscow State University, Moscow, 119991 Russia*
[b] *Moscow Technical University of Communications and Informatics, Moscow, 123423 Russia*
[c] *Dubna State University, Moscow region, Moscow, 141980 Russia*
[d] *Joint Institute for Nuclear Research, Moscow region, Moscow, 141980 Russia*

**Abstract**

On the basis of the Vlasov chain of equations, a new infinite dispersion chain of equations is obtained for the distribution functions of mixed higher order kinematical values. In contrast to the Vlasov chain, the dispersion chain contains distribution functions with an arbitrary set of kinematical values and has a tensor form of writing. For the dispersion chain, new equations for mixed Boltzmann functions and the corresponding chain of conservation laws for fluid dynamics are obtained. The probability is proved to be a constant value for a particle to belong the region where the quasi-probability density is negative (Wigner function).

**Introduction**

An infinite chain of self-linking Vlasov equations is written for the distribution functions of kinematical values of all orders $f_1(\vec{r},t)$, $f_2(\vec{r},\vec{v},t)$, $f_3(\vec{r},\vec{v},\dot{\vec{v}},t)$, $f_4(\vec{r},\vec{v},\dot{\vec{v}},\ddot{\vec{v}},t),\ldots$ The variables $\vec{r},\vec{v},\dot{\vec{v}},\ddot{\vec{v}},\ldots$ are independent values. It is important to note that the Vlasov chain is derived from the first principle – the conservation law for probabilities in a generalized phase space containing the entire set of kinematical values $\vec{r},\vec{v},\dot{\vec{v}},\ddot{\vec{v}},\ldots$ A. Vlasov in his monograph [1] noted an important difference of the description he proposed: «*The equation for the distribution function $f(\vec{r},\vec{p},t)$ was obtained in a structure close to the Liouville equation. But this name is not accurate, since the conservation law for probabilities has a different (not mechanical) nature:* $\dim f(\vec{q}_i(t),\vec{p}_i(t),t)_{\text{Liuv}} = 1$, $\dim f(\vec{q}_i,\vec{p}_i,t)_{\text{stat}} = 6N+1$. *For no number $N$ of particles does the topological dimension of the statistical distribution function coincide with the invariants of the Liouville equation*». Vlasov demonstrated that the equations of the chain might be represented as classical mechanics equations when «shrinking» the distribution $f$ to the $\delta-$function. He formulated this assertion in the form of the following theorem [1].

<u>***Theorem***</u>. *The transition from statistics to particle mechanics is possible in case of*
1) *lowering the topological dimension of mechanical elements in statistics,*
2) *elimination of the statistical spread of coordinates, velocities and accelerations,*
3) *conservation of the statistical element and the limiting transition to mechanics (or in other words normalization of the distribution functions).*

The Vlasov chain cut-off leads, on the one hand, to the loss of information about the system, and on the other hand, to the introduction of dynamic approximations of the mean kinematical values. For instance, for the first equation in the Vlasov chain, a cut-off leads to a certain type of motion equation (velocity is proportional to force $\vec{F}$):

$$\langle \vec{v} \rangle = \frac{1}{\kappa m} \vec{F_1}, \tag{i.1}$$

where $\frac{1}{\kappa m}$ is an empirical proportionality coefficient. It is known that this law of motion (i.1) belongs to Aristotle. According to the Helmholtz theorem expression (i.1) can be represented in the form [2]

$$\langle \vec{v} \rangle (\vec{r},t) = -\alpha \nabla_r \Phi(\vec{r},t) + \gamma \vec{A}(\vec{r},t), \tag{i.2}$$

where $\vec{A}$ corresponds to the vector potential of the magnetic field $\vec{B} = \operatorname{rot} \vec{A}$, and $\Phi = 2\varphi + 2\pi k$, $k \in \mathbb{Z}$ is in fact the phase of the wave function $\Psi(\vec{r},t) = |\Psi| e^{i\varphi}$ and satisfies the Hamilton-Jacobi equation:

$$-\hbar \frac{\partial \varphi}{\partial t} = \frac{m}{2} |\langle \vec{v} \rangle|^2 + e\chi = \mathrm{H}, \tag{i.3}$$

$$e\chi \stackrel{\text{det}}{=} U + \mathrm{Q} + \frac{e^2}{2m} |\vec{A}|^2, \quad \mathrm{Q} = \frac{\alpha}{\beta} \frac{\Delta |\Psi|}{|\Psi|} = -\frac{\hbar^2}{2m} \frac{\Delta |\Psi|}{|\Psi|},$$

where $\alpha \stackrel{\text{det}}{=} -\frac{\hbar}{2m}$, $\beta \stackrel{\text{det}}{=} \frac{1}{\hbar}$, $\gamma \stackrel{\text{det}}{=} -\frac{e}{m}$. Value Q is the quantum potential from the de Broglie-Bohm theory of the «pilot wave» [3-5]. Note that potential $e\chi$ (i.3) in classical mechanics (at $\hbar \to 0$) and in the absence of a vortex field ($\vec{A} = \vec{0}$) goes over into potential $U$.

Hamiltonian H (i.3) is related to Lagrange function L through the Legendre transform

$$\mathrm{L} + \mathrm{H} = m(\langle \vec{v} \rangle, \langle \vec{v}_p \rangle), \tag{i.4}$$

where $\langle \vec{v}_p \rangle = -\alpha \nabla \Phi$ is the vortex-free (or conservative) component of the vector field of the probability flux (i.2).

The distribution function $f_1(\vec{r},t) = |\Psi(\vec{r},t)|^2 \geq 0$ determines the amplitude of the wave function $\Psi = \sqrt{f_1} e^{i\varphi}$ and satisfies the Schrödinger equation:

$$\frac{i}{\beta} \frac{\partial \Psi}{\partial t} = -\alpha \beta \left( \hat{\mathrm{p}} - \frac{\gamma}{2\alpha\beta} \vec{A} \right)^2 \Psi + V\Psi, \tag{i.5}$$

$$V \stackrel{\text{det}}{=} \frac{1}{2\alpha\beta} \frac{|\gamma \vec{A}|^2}{2} + U, \quad \hat{\mathrm{p}} \stackrel{\text{det}}{=} -\frac{i}{\beta} \nabla, \quad \hat{\mathrm{p}}^2 = -\frac{1}{\beta^2} \Delta.$$

The cut-off in the second equation of the Vlasov chain leads to the equation of motion:

$$\langle \dot{\vec{v}} \rangle = \vec{F_2}(\vec{r},\vec{v},t), \tag{i.6}$$

which is similar to the equation of motion of the Newtonian mechanics. Note that expression (i.6), on the one hand, may be written in the form of hydrodynamic equation [1, 6]:

$$\frac{d\langle v_\mu \rangle}{dt} = \left( \frac{\partial}{\partial t} + \langle v_\lambda \rangle \frac{\partial}{\partial x_\lambda} \right) \langle v_\mu \rangle = -\frac{1}{f_1} \frac{\partial P_{\mu\lambda}}{\partial x_\lambda} + \langle \langle \dot{v}_\mu \rangle \rangle, \qquad (i.7)$$

where $P_{\mu\lambda} = \int_{(\infty)} f_2 (v_\mu - \langle v_\mu \rangle)(v_\lambda - \langle v_\lambda \rangle) d^3 v$ is a pressure tensor, which, from the standpoint of quantum mechanics, determines the quantum pressure tensor $P_{\mu\lambda}^{(q)}$ and is related to the quantum potential Q (i.3):

$$-\frac{1}{f_1} \frac{\partial P_{\mu\lambda}^{(q)}}{\partial x^\lambda} = 2\alpha^2 \frac{\partial}{\partial x^\mu} \left( \frac{1}{\sqrt{f_1}} \frac{\partial^2 \sqrt{f_1}}{\partial x^\lambda \partial x^\lambda} \right) = 2\alpha\beta \frac{\partial Q}{\partial x^\mu}. \qquad (i.8)$$

On the other hand, expression (i.6) admits the Vlasov-Moyal approximation [7, 8] and may be represented as

$$\langle \dot{v}_\mu \rangle = \sum_{n=0}^{+\infty} \frac{(-1)^{n+1}(\hbar/2)^{2n}}{m^{2n+1}(2n+1)!} \frac{\partial^{2n+1} U}{\partial x_\mu^{2n+1}} \frac{1}{f_2} \frac{\partial^{2n} f_2}{\partial v_\mu^{2n}}, \qquad (i.9)$$

where $\langle \langle \dot{v}_\mu \rangle \rangle = -\frac{1}{m} \frac{\partial U}{\partial x_\mu}$, which, being substituted into equation (i.7), gives an approximation of an external force with potential $U$. The distribution function $f_2(\vec{r}, \vec{v}, t)$ in this case is the known Wigner function [9], which is used to describe quantum systems in a phase space. Note that equation (i.7) in accordance with expressions (i.2), (i.3) may be written in the electromagnetic form [2]:

$$\frac{d}{dt} \langle \vec{v} \rangle = -\gamma \left( \vec{E} + \langle \vec{v} \rangle \times \vec{B} \right), \qquad (i.10)$$

$$\vec{E} = -\frac{\partial}{\partial t} \vec{A} - \nabla \chi.$$

The Vlasov chain cut-off on the third equation gives the following representation:

$$\langle \ddot{\vec{v}} \rangle = \vec{F}_3 \left( \vec{r}, \vec{v}, \dot{\vec{v}}, t \right). \qquad (i.11)$$

Equation (i.11) may be used when considering the motion of a charged particle with acceleration (Lorentz equation) [10]:

$$m\dot{\vec{v}} = \vec{F}_{ext} + \frac{e^2}{6\pi\varepsilon_0 c^3} \ddot{\vec{v}}, \qquad (i.12)$$

where $\vec{F}_{ext}$ is an external force. Derivative $\ddot{\vec{v}}$ contains information about the radiation reaction force and determines the third order of the differential motion equation (i.12), which is beyond classical mechanics. Note that particle mechanics does not answer the question why the equation of motion is one of the second order as they are initial equations in this theory.

It can be seen that the higher the equation number in the Vlasov chain cut-off the larger amount of the information we can get about a physical system: Aristotle's law, Newton's, Lorentz's equations. The equation number at which we cut-off the chain depends on how much information of the system we need. The Vlasov chain describes a fundamental approach to such «information-step» consideration, not limited to phenomenology. In a natural way, the Vlasov chain of equations links classical and statistical mechanics, field theory, continuum mechanics, plasma physics, astrophysics, quantum mechanics, and accelerator physics [11-15].

Due to the fundamental importance of the Vlasov equations chain, the aim of this work is to extend the initial set of the Vlasov equations with distribution functions of «mixed» type. The problem is as follows. The Vlasov equations chain has a hierarchical structure, that is, the equations are considered for the functions $f_1(\vec{r},t)$, $f_2(\vec{r},\vec{v},t)$, $f_3(\vec{r},\vec{v},\dot{\vec{v}},t)$, $f_4(\vec{r},\vec{v},\dot{\vec{v}},\ddot{\vec{v}},t),\ldots$:

$$\frac{\partial f_1}{\partial t} + \text{div}_r\left[f_1 \langle \vec{v} \rangle\right] = 0,$$

$$\frac{\partial f_2}{\partial t} + \text{div}_r\left[f_2 \vec{v}\right] + \text{div}_v\left[f_2 \langle \dot{\vec{v}} \rangle\right] = 0,$$

$$\frac{\partial f_3}{\partial t} + \text{div}_r\left[f_3 \vec{v}\right] + \text{div}_v\left[f_3 \dot{\vec{v}}\right] + \text{div}_{\dot{v}}\left[f_3 \langle \ddot{\vec{v}} \rangle\right] = 0,$$

$$\ldots$$

$$\frac{\partial f_n}{\partial t} + \text{div}_r\left[f_n \vec{v}\right] + \text{div}_v\left[f_n \dot{\vec{v}}\right] + \ldots + \text{div}_{\overset{(n-2)}{v}} f_n \left\langle \overset{(n-1)}{\vec{v}} \right\rangle = 0,$$

$$\ldots$$

(i.13)

The distribution functions are related to each other by the conditions:

$$f_0(t) = \int_{(\infty)} f_1(\vec{r},t) d^3r = \int_{(\infty)} f_2(\vec{r},\vec{v},t) d^3v =$$

$$= \int_{(\infty)}\int_{(\infty)} f_3(\vec{r},\vec{v},\dot{\vec{v}},t) d^3v d^3\dot{v} = \int_{(\infty)}\int_{(\infty)}\int_{(\infty)} f_4(\vec{r},\vec{v},\dot{\vec{v}},\ddot{\vec{v}},t) d^3v d^3\dot{v} d^3\ddot{v} = \ldots,$$

(i.14)

The value $f_0(t)$ may correspond to the number of particles in the system $N(t)$ or be a normalization factor. Mean kinematical values $\langle \vec{v} \rangle$, $\langle \dot{\vec{v}} \rangle$, $\langle \ddot{\vec{v}} \rangle$,… in chain (i.13) are determined by the relations

$$f_1(\vec{r},t)\langle \vec{v} \rangle(\vec{r},t) = \int_{(\infty)} f_2(\vec{r},\vec{v},t)\vec{v} d^3v,$$

$$f_2(\vec{r},\vec{v},t)\langle \dot{\vec{v}} \rangle(\vec{r},\vec{v},t) = \int_{(\infty)} f_3(\vec{r},\vec{v},\dot{\vec{v}},t)\dot{\vec{v}} d^3\dot{v},$$

$$f_3(\vec{r},\vec{v},\dot{\vec{v}},t)\langle \ddot{\vec{v}} \rangle(\vec{r},\vec{v},\dot{\vec{v}},t) = \int_{(\infty)} f_4(\vec{r},\vec{v},\dot{\vec{v}},\ddot{\vec{v}},t)\ddot{\vec{v}} d^3\ddot{v},$$

(i.15)

$$\ldots$$

Note that relations (i.15) admit additional averaging:

$$f_0(t)\langle\langle \vec{v} \rangle\rangle(t) = \int_{(\infty)} f_1(\vec{r},t)\langle \vec{v} \rangle(\vec{r},t) d^3r,$$

$$f_1(\vec{r},t)\langle\langle\dot{\vec{v}}\rangle\rangle(\vec{r},t) = \int_{(\infty)} f_2(\vec{r},\vec{v},t)\langle\dot{\vec{v}}\rangle(\vec{r},\vec{v},t)d^3v, \qquad (i.16)$$

$$f_0(t)\langle\langle\langle\dot{\vec{v}}\rangle\rangle\rangle(t) = \int_{(\infty)} f_1(\vec{r},t)\langle\langle\dot{\vec{v}}\rangle\rangle(\vec{r},t)d^3r,$$

...

Since kinematical values $\vec{r}, \vec{v}, \dot{\vec{v}}, \ddot{\vec{v}},...$ are independent ones, it seems logical to have information on distribution functions of the «mixed» type, for instance,

$$f(\vec{v},t),\ f(\dot{\vec{v}},t),\ f(\ddot{\vec{v}},t),... \qquad (i.17)$$

$$f(\vec{r},\dot{\vec{v}},t),\ f(\vec{r},\ddot{\vec{v}},t),....f(\vec{v},\dot{\vec{v}},t),f(\vec{v},\ddot{\vec{v}},t),...,f(\dot{\vec{v}},\ddot{\vec{v}},t),f(\dot{\vec{v}},\dddot{\vec{v}},t),...$$

$$f(\vec{r},\vec{v},\ddot{\vec{v}},t),...,f(\vec{r},\dot{\vec{v}},\ddot{\vec{v}},t),...$$

as well as on mixed mean kinematical values

$$\langle\dot{\vec{v}}\rangle(\vec{v},t),\ \langle\vec{v}\rangle(\dot{\vec{v}},t),\ \langle\vec{v}\rangle(\vec{r},\dot{\vec{v}},t),\ \langle\dot{\vec{v}}\rangle(\vec{r},\vec{v},\ddot{\vec{v}},t),... \qquad (i.18)$$

Note that mean kinematical values in the Vlasov equations are functions depending on the kinematical values of the lowest order, that is $\langle\vec{v}\rangle(\vec{r},t)$, $\langle\dot{\vec{v}}\rangle(\vec{r},\vec{v},t)$, $\langle\ddot{\vec{v}}\rangle(\vec{r},\vec{v},\dot{\vec{v}},t)$,... or, after further averaging, $\langle\langle\dot{\vec{v}}\rangle\rangle(\vec{r},t)$, $\langle\langle\dot{\vec{v}}\rangle\rangle(\vec{v},t)$, $\langle\langle\ddot{\vec{v}}\rangle\rangle(\vec{r},\vec{v},t)$, $\langle\langle\langle\ddot{\vec{v}}\rangle\rangle\rangle(\vec{r},t)$. The consideration of equations (i.7) or (i.12) may be carried out in different ways. On the one hand, looking at equation (i.7), one can assume that value $\dot{\vec{v}}$ is determined by value $\vec{v}$, that is, the highest kinematical value depends on the lowest kinematical value in accordance with the representation (i.6). Similarly, in equation (i.12), one can assume that value $\ddot{\vec{v}}$ is determined by $\dot{\vec{v}}$ in accordance with representation (i.11). Such reasoning is explained by the hierarchical form of writing the Vlasov chain of equations (i.13). On the other hand, due to the independence of kinematical values $\vec{r}, \vec{v}, \dot{\vec{v}}, \ddot{\vec{v}},...$, it is possible to construct mixed distribution functions (i.17) and the corresponding average kinematical values (i.18). In this case, equation (i.12) may be interpreted as the dependence of the lowest kinematical value $\dot{\vec{v}}$ on the highest kinematical value $\ddot{\vec{v}}$. Similar reasoning may be carried out for equation (i.7). Apparently, this approach is close to classical mechanics, if one recalls the expansion in the Taylor series of the trajectory of a material point $\vec{r}(t) = \vec{r}_0 + \vec{v}_0 t + \dot{\vec{v}}_0 \frac{t^2}{2} + \ddot{\vec{v}}_0 \frac{t^3}{3!} + ...$, similarly for $\vec{v}(t) = \vec{v}_0 + \dot{\vec{v}}_0 t + \ddot{\vec{v}}_0 \frac{t^2}{2!} + ...$ and further.

In this work, an infinite chain is constructed for the distribution functions of mixed type (i.17) and the kinematical values corresponding to them (i.18), which was called the *dispersion chain of the Vlasov equations*. By analogy with work [6], groups of equations for mixed $H_n$-Boltzmann functions are constructed and their properties are investigated. Dispersion equations of conservation laws for mixed distribution functions are obtained as well.

The work has the following structure. In § 1 the work describes the general formalism for representing distribution functions and kinematical value of the mixed type. It is based on the so-called «extensive» notion used in tensor analysis. The use of such an extensive of various ranks to represent distribution functions, high order kinematical values, differential operators allows one to obviously analyze the properties of these objects and greatly simplifies mathematical transforms. In § 2, using the formalism of the extensive, we define the dispersion chain of the Vlasov equations. The shape of the dispersion chain clearly shows the change in the distribution function along the phase trajectory due to sources of dissipation $Q$ determined by higher kinematical values. In §3, equations of conservation laws are obtained for higher kinematical values. In the simplest classical case (only the coordinate representation), these laws correspond to the law of conservation of mass, momentum and energy. A number of theorems on the form of motion equations with higher kinematical value is proved. In § 4, the concept of the $H^{n_1,...,n_R}$ -Boltzmann function is introduced, which is a generalization of the known $H$ -Boltzmann function for a generalized phase space of higher kinematical values. In a particular case, $H^{1,2}$ – function coincides with $H$ -Boltzmann function for a phase space $\{\vec{r}, m\vec{v}\}$. The theorem is proved that the evolution of $H^{n_1,...,n_R}$ -function is determined by the signs of dissipation sources $Q$. The mean values of the functions $\langle Q \rangle$ are the sources of entropy production in a generalized phase space. For quantum mechanics in a phase space, the case of a distribution function with negative probabilities (Wigner function) is considered. The theorem is proved that the area of the phase region with negative values of the probabilities remains unchanged over time. The Conclusion section presents the main results of the work. The Appendix contains theorem proofs and intermediate mathematical manipulations.

## §1 Extensive form

When analyzing possible set of distribution functions (i.17), it is seen that it is possible to rank them. Function $f_0(t)$ (i.14) will have the smallest zero rank. Integration of all distribution functions over entire infinite-dimensional generalized phase space $\Omega$ [6, 16] will lead to one single function $f_0(t)$, which will also be denoted as $f^0(t)$. Let us consider an ordered infinite set of distribution functions of the 1st rank $F^1$:

$$F^1 \mapsto \{f^j\} = \left( f^1(\vec{r},t) \quad f^2(\vec{v},t) \quad f^3(\dot{\vec{v}},t) \quad ... \right), \qquad (1.1)$$

where superscript «i» corresponds to the number of the kinematical value: $\vec{r}, \vec{v}, \dot{\vec{v}}, \ddot{\vec{v}},...$

The distribution functions of the 2nd rank form set $F^2$, which we represent in the form of an infinite matrix:

$$F^2 \mapsto \{f^{i,j}\} = \begin{pmatrix} \times & f^{1,2}(\vec{r},\vec{v},t) & f^{1,3}(\vec{r},\dot{\vec{v}},t) & f^{1,4}(\vec{r},\ddot{\vec{v}},t) & ... \\ f^{1,2}(\vec{r},\vec{v},t) & \times & f^{2,3}(\vec{v},\dot{\vec{v}},t) & f^{2,4}(\vec{v},\ddot{\vec{v}},t) & ... \\ f^{1,3}(\vec{r},\dot{\vec{v}},t) & f^{2,3}(\vec{v},\dot{\vec{v}},t) & \times & f^{3,4}(\dot{\vec{v}},\ddot{\vec{v}},t) & ... \\ f^{1,4}(\vec{r},\ddot{\vec{v}},t) & f^{2,4}(\vec{v},\ddot{\vec{v}},t) & f^{3,4}(\dot{\vec{v}},\ddot{\vec{v}},t) & \times & ... \\ ... & ... & ... & ... & ... \end{pmatrix}, \qquad (1.2)$$

where the symmetry condition $f^{n_1,n_2} = f^{n_2,n_1}$, $n_1, n_2 \in \mathbb{N}$ is taken into account. The diagonal of matrix (1.2) is empty, since corresponding elements $f^{n,n} = \{f^{1,1}(\vec{r},\vec{r},t), f^{2,2}(\vec{v},\vec{v},t),...\}$ are the functions of one kinematical value (1.1). Instead of «×» sign in matrix (1.2), we will sometimes use the number zero.

A similar construction can be performed for sets («extensives») $F^R$ of distribution functions of arbitrary rank $R$. For the sake of compactness, we will use the $\Psi$-algebra formalism [16], using the following notation:

$$\vec{\xi}^4 = \ddot{\vec{v}}, \quad \vec{\xi}^{3,6} = \left(\dot{\vec{v}}, \dddot{\vec{v}}\right)^T, \quad \vec{\xi}^{1,2,3,4} = \left(\vec{r}, \vec{v}, \dot{\vec{v}}, \ddot{\vec{v}}\right)^T, \ldots \quad (1.3)$$

$$f^{1,2}(\vec{r}, \vec{v}, t) = f^{1,2}\left(\vec{\xi}^{1,2}, t\right) = f^{1,2}, \ldots$$

$$f^{n,\ldots,n} \stackrel{\text{det}}{=} f_n, \quad n \in \mathbb{N}.$$

When writing distribution function $f^{n_1,\ldots,n_R}$ of rank $R$, we will assume that the sequence of indices $n_1,\ldots,n_R$ is increasing and all indices are unique, that is $n_1 \neq \ldots \neq n_R$. Distribution function $f^{n_1,\ldots,n_R}$ is defined in subspace $\Omega^{n_1} \times \cdots \times \Omega^{n_R}$ of infinite-dimensional generalized phase space $\Omega$ [16], containing a complete set of kinematical values of all orders. The set of $f^{n_1,\ldots,n_R}$ functions forms an extensive $F^R$ of rank $R$.

In addition to rank $R$, we will use the notion of a set $S$ of arguments of distribution function $f^{n_1,\ldots,n_R}$ of the extensive $F^{R,S}$. Set $S$ is a set of kinematical values of the form: $S = \{\ddot{\vec{v}}\}$, $S = \{\dot{\vec{v}}, \dddot{\vec{v}}\}$, $S = \{\vec{r}, \vec{v}, \dot{\vec{v}}, \ddot{\vec{v}}\}$, etc. For the sake of compactness of writing the elements of set $S$, we will also use the representation $S = \{4\}$, $S = \{3,6\}$, $S = \{1,2,3,4\}$.

Let there be some arbitrary fixed set of kinematical values $S = \{s_1,\ldots,s_R\}$. Set $S$ determines distribution function $f^{s_1,\ldots,s_R}$ of rank $R = |S|$. Corresponding extension $F^{R,S}$ will only have one element $f^{s_1,\ldots,s_R}$. The extensions of ranks $R-1, R-2, \ldots, 1$ will have $C_{|S|}^{R-1}, C_{|S|}^{R-2}, \ldots, C_{|S|}^1$ elements respectively. For instance, for set $S = \{\vec{r}, \dot{\vec{v}}, \dddot{\vec{v}}\}$ and ranks $R=1$ and $R=2$ extensions $F^{1,S}$ and $F^{2,S}$ have $C_3^1 = C_3^2 = 3$ elements, and extension $F^{3,S}$ has only one element:

$$F^{1,\{1,3,5\}} = \left\{f^1, \times, f^3, \times, f^5\right\}, \quad F^{2,\{1,3,5\}} = \begin{pmatrix} \times & \times & f^{1,3} & \times & f^{1,5} \\ \times & \times & \times & \times & \times \\ f^{1,3} & \times & \times & \times & f^{3,5} \\ \times & \times & \times & \times & \times \\ f^{1,5} & \times & f^{3,5} & \times & \times \end{pmatrix}, \quad F^{3,\{1,3,5\}} = \left\{f^{1,3,5}\right\}. \quad (1.4)$$

By analogy with expressions (i.14), the relation between distribution functions of different ranks is defined as follows.

**Definition 1.** *Let $k = \{k_1,\ldots,k_K\}$, $k_1 \neq \ldots \neq k_K$ subset of set $n = \{n_1,\ldots,n_R\}$, $n_1 \neq \ldots \neq n_R$, $K \leq R$, then we define the distribution function of $R-K$ rank as*

$$f^{n\setminus k}\left(\vec{\xi}^{n\setminus k}, t\right) \stackrel{\text{det}}{=} \int_{\Omega^{k_1}} \cdots \int_{\Omega^{k_K}} f^n\left(\vec{\xi}^n, t\right) \prod_{s=1}^K d^3\xi^{k_s}. \quad (1.5)$$

*where $n\setminus k$ is the difference of sets $n$ and $k$.*

**Remark.** For a fixed set $n = \{n_1, n_2, ..., n_R\}$ and various collections of sets $k = \{k_1, k_2, ..., k_K\}$ such that $k \subset n$ one can obtain $C_R^K$ different distribution functions of $R-K$ rank. As an example, let us consider function $f^{1,3,4,6}$, for which $R = 4$, $n = \{1,3,4,6\}$. At $K = 1$ there exist $C_4^1 = 4$ collections of sets $n \setminus k$:

$$(n \setminus k)^1 = \{1,3,4\}, \, (n \setminus k)^2 = \{1,3,6\}, \, (n \setminus k)^3 = \{1,4,6\}, \, (n \setminus k)^4 = \{3,4,6\}. \tag{1.6}$$

Each set (1.6) corresponds to the distribution function of the third rank ($R - K = 3$):

$$\begin{aligned}
f^{(n\setminus k)^1} &= f^{1,3,4}\left(\vec{\xi}^{1,3,4}, t\right) = f^{1,3,4}\left(\vec{r}, \dot{\vec{v}}, \ddot{\vec{v}}, t\right) = \int_{\Omega^6} f^{1,3,4,6} d^3\dddot{v}, \\
f^{(n\setminus k)^2} &= f^{1,3,6}\left(\vec{\xi}^{1,3,6}\right) = f^{1,3,6}\left(\vec{r}, \dot{\vec{v}}, \dddot{\vec{v}}, t\right) = \int_{\Omega^4} f^{1,3,4,6} d^3\ddot{v}, \\
f^{(n\setminus k)^3} &= f^{1,4,6}\left(\vec{\xi}^{1,4,6}\right) = f^{1,4,6}\left(\vec{r}, \ddot{\vec{v}}, \dddot{\vec{v}}, t\right) = \int_{\Omega^3} f^{1,3,4,6} d^3\dot{v}, \\
f^{(n\setminus k)^4} &= f^{3,4,6}\left(\vec{\xi}^{3,4,6}\right) = f^{3,4,6}\left(\dot{\vec{v}}, \ddot{\vec{v}}, \dddot{\vec{v}}, t\right) = \int_{\Omega^1} f^{1,3,4,6} d^3r.
\end{aligned} \tag{1.7}$$

Functions (1.7) are obtained from function $f^n = f^{1,3,4,6}\left(\vec{\xi}^{1,3,4,6}, t\right)$ by integration over subspaces $\Omega^6$, $\Omega^4$, $\Omega^3$, $\Omega^1$ respectively. If a repeated integration of distribution functions (1.7) is performed then we obtain distribution functions of the 1st and 2nd rank. Indeed, for the distribution functions of the 2nd rank, we obtain a set of $C_3^1 = 3$ sets $m = \{m_1, m_2\}$ for each set $(n \setminus k)^j$, $j = 1...4$ (1.6):

$$\begin{aligned}
(n \setminus k)^1 &\mapsto m^1 = \{1,3\}, \, m^2 = \{1,4\}, \, m^3 = \{3,4\}, \\
(n \setminus k)^2 &\mapsto m^1 = \{1,3\}, \, m^2 = \{1,6\}, \, m^3 = \{3,6\}, \\
(n \setminus k)^3 &\mapsto m^1 = \{1,4\}, \, m^2 = \{1,6\}, \, m^3 = \{4,6\}, \\
(n \setminus k)^4 &\mapsto m^1 = \{3,4\}, \, m^2 = \{3,6\}, \, m^3 = \{4,6\}.
\end{aligned} \tag{1.8}$$

Eliminating repeating variants from expressions (1.8), we obtain $C_4^2 = 6$ variants of distribution functions of the 2nd rank ($R - K = 2$)

$$\begin{pmatrix}
\times & \times & f^{1,3} & f^{1,4} & \times & f^{1,6} \\
\times & \times & \times & \times & \times & \times \\
f^{1,3} & \times & \times & f^{3,4} & \times & f^{3,6} \\
f^{1,4} & \times & f^{3,4} & \times & \times & f^{4,6} \\
\times & \times & \times & \times & \times & \times \\
f^{1,6} & \times & f^{3,6} & f^{4,6} & \times & \times
\end{pmatrix} \tag{1.9}$$

There will be $C_4^3 = 4$ kinds of distribution functions of the 1st ranks ($N - K = 1$)

$$f^1(\vec{r},t),\ f^3(\dot{\vec{v}},t),\ f^4(\ddot{\vec{v}},t),\ f^6(\dddot{\vec{v}},t). \tag{1.10}$$

From representations (1.7), (1.9), (1.10) it is seen that not all possible distribution functions of the 3$^{rd}$, 2$^{nd}$ and 1$^{st}$ rank can be found from function $f^{1,3,4,6}$.

Let us consider a set of mean kinematical values (i.14), (i.15), (i.17).

**Definition 2.** Let $k=\{k_1,...,k_K\}$, $k_1 \neq ... \neq k_K$ is a subset of set $n=\{n_1,...,n_R\}$, $n_1 \neq ... \neq n_R \neq \ell \in \mathbb{N}$, then we define mean kinematical values of $\ell$ order and rank $R-K$ as

$$\left\langle \vec{\xi}^\ell \right\rangle_{n\backslash k} = \frac{1}{f^{n\backslash k}} \int_{\Omega^{k_1}} ... \int_{\Omega^{k_K}} \int_{\Omega^\ell} f^{n\cup\{\ell\}}\left(\vec{\xi}^{n\cup\{\ell\}},t\right)\vec{\xi}^\ell d^3\xi^\ell \prod_{s=1}^{K} d^3\xi^{k_s}, \tag{1.11}$$

*where $n\cup\{\ell\}$ is a union of set $n$ and element $\ell$, and $n\backslash k$ is the difference of sets $n$ and $k$. Distribution function $f^{n\cup\{\ell\}}$ is of rank $R+1$.*

For fixed set $S$, a set of mean kinematical values of order $\ell \notin S$ and rank $R$ form an extensive $\vec{\Xi}^\ell_{R,S}$. For instance, for set $S=\{1,3,4,6\}$ and rank $R=2$, we obtain

$$\vec{\Xi}^\ell_{2,\{1,3,4,6\}} = \begin{pmatrix} \times & \times & \left\langle \vec{\xi}^\ell \right\rangle_{1,3} & \left\langle \vec{\xi}^\ell \right\rangle_{1,4} & \times & \left\langle \vec{\xi}^\ell \right\rangle_{1,6} \\ \times & \times & \times & \times & \times & \times \\ \left\langle \vec{\xi}^\ell \right\rangle_{1,3} & \times & \times & \left\langle \vec{\xi}^\ell \right\rangle_{3,4} & \times & \left\langle \vec{\xi}^\ell \right\rangle_{3,6} \\ \left\langle \vec{\xi}^\ell \right\rangle_{1,4} & \times & \left\langle \vec{\xi}^\ell \right\rangle_{3,4} & \times & \times & \left\langle \vec{\xi}^\ell \right\rangle_{4,6} \\ \times & \times & \times & \times & \times & \times \\ \left\langle \vec{\xi}^\ell \right\rangle_{1,6} & \times & \left\langle \vec{\xi}^\ell \right\rangle_{3,6} & \left\langle \vec{\xi}^\ell \right\rangle_{4,6} & \times & \times \end{pmatrix}. \tag{1.12}$$

By analogy with expression (1.9), extensive (1.12) may be obtained from mean kinematical value $\left\langle \vec{\xi}^\ell \right\rangle_{1,3,4,6}$ by its averaging (1.11) over subspaces $\Omega^6$, $\Omega^4$, $\Omega^3$, $\Omega^1$.

**§2 Dispersion chain**

Based on the Vlasov equations chain (i.13), we obtain equations that are satisfied by the mixed distribution functions (1.5). Integration of the second equation in chain (i.13) over the velocity space will give the first equation for function $f^1$. We integrate the second equation over the coordinate space and obtain an equation for function $f^2$

$$\frac{\partial}{\partial t}\int_{(\infty)} f^{1,2} d^3r + \int_{(\infty)} \text{div}_r\left[f^{1,2}\vec{v}\right]d^3r + \text{div}_v \int_{(\infty)} f^{1,2}\left\langle\dot{\vec{v}}\right\rangle_{1,2} d^3r = 0,$$

$$\frac{\partial}{\partial t}f^2 + \text{div}_v\left[f^2\left\langle\dot{\vec{v}}\right\rangle_2\right] = 0, \tag{2.1}$$

where the integration uses the Ostrogradsky-Gauss theorem and the condition for the fast decrease of the distribution function at infinity [1].

Let us consider the third equation in chain (i.13) for function $f^{1,2,3}$. Integrating the third equation over the spaces of coordinates and accelerations we obtain equation (2.1). The equation for function $f^{1,2,3}$ is obtained through integration over the spaces of coordinates and velocities

$$\frac{\partial}{\partial t}\int_{(\infty)}\int_{(\infty)} f^{1,2,3} d^3r d^3v + \int_{(\infty)}\int_{(\infty)} \operatorname{div}_r\left[f^{1,2,3}\vec{v}\right]d^3r d^3v + \int_{(\infty)}\int_{(\infty)} \operatorname{div}_v\left[f^{1,2,3}\dot{\vec{v}}\right]d^3r d^3v +$$

$$+ \int_{(\infty)}\int_{(\infty)} \operatorname{div}_{\dot{v}}\left[f^{1,2,3}\left\langle\ddot{\vec{v}}\right\rangle_{1,2,3}\right]d^3r d^3v = 0,$$

$$\frac{\partial}{\partial t}f^3 + \operatorname{div}_{\dot{v}}\left[f^3\left\langle\ddot{\vec{v}}\right\rangle_3\right] = 0. \tag{2.2}$$

We obtain equations for distribution functions $f^{2,3}$ and $f^{1,3}$:

$$\frac{\partial}{\partial t}\int_{(\infty)} f^{1,2,3}d^3r + \int_{(\infty)} \operatorname{div}_r\left[f^{1,2,3}\vec{v}\right]d^3r + \operatorname{div}_v\left[\int_{(\infty)} f^{1,2,3}\dot{\vec{v}}d^3r\right] + \int_{(\infty)} \operatorname{div}_{\dot{v}}\left[f^{1,2,3}\left\langle\ddot{\vec{v}}\right\rangle_{1,2,3}\right]d^3r = 0,$$

$$\frac{\partial}{\partial t}f^{2,3} + \operatorname{div}_v\left[f^{2,3}\dot{\vec{v}}\right] + \operatorname{div}_{\dot{v}}\left[f^{2,3}\left\langle\ddot{\vec{v}}\right\rangle_{2,3}\right] = 0, \tag{2.3}$$

and

$$\frac{\partial}{\partial t}\int_{(\infty)} f^{1,2,3}d^3v + \int_{(\infty)} \operatorname{div}_r\left[f^{1,2,3}\vec{v}\right]d^3v + \int_{(\infty)} \operatorname{div}_v\left[f^{1,2,3}\dot{\vec{v}}\right]d^3v + \int_{(\infty)} \operatorname{div}_{\dot{v}}\left[f^{1,2,3}\left\langle\ddot{\vec{v}}\right\rangle_{1,2,3}\right]d^3v = 0,$$

$$\frac{\partial}{\partial t}f^{1,3} + \operatorname{div}_r\left[f^{1,3}\left\langle\vec{v}\right\rangle_{1,3}\right] + \operatorname{div}_{\dot{v}}\left[f^{1,3}\left\langle\ddot{\vec{v}}\right\rangle_{1,3}\right] = 0. \tag{2.4}$$

Note that equation (2.3) is similar in form with the second equation in chain (i.13), and equation (2.4) is substantially differs from all the equations in the chain.

Let us analyze the fourth equation in chain (i.13). As noted in §1, distribution function $f^{1,2,3,4}$ of the fourth rank should yield $C_4^1 = 4$ functions of the first rank, $C_4^2 = 6$ functions of the second rank and $C_4^3 = 4$ functions of the third rank. The equations for distribution function $f^1$, $f^2$, $f^3$ of the first rank were obtained earlier (i.13), (2.1), (2.2). Functions $f^{1,2}$, $f^{2,3}$, $f^{1,3}$ of the second rank satisfy equations (i.13), (2.3), (2.4) respectively. The equation for function $f^{1,2,3}$ of the third rank is the third equation in chain (i.13). Thus, by integrating the Vlasov chain over the remaining combinations of kinematical subspaces, we obtain new equations for mixed functions: $f^4$, $f^{1,4}$, $f^{2,4}$, $f^{3,4}$, $f^{1,2,4}$, $f^{2,3,4}$, $f^{1,3,4}$

$$\frac{\partial f^4}{\partial t} + \operatorname{div}_{\ddot{v}}\left[f^4\left\langle\dddot{\vec{v}}\right\rangle_4\right] = 0, \tag{2.5}$$

$$\frac{\partial f^{1,4}}{\partial t} + \operatorname{div}_r\left[f^{1,4}\left\langle\vec{v}\right\rangle_{1,4}\right] + \operatorname{div}_{\ddot{v}}\left[f^{1,4}\left\langle\dddot{\vec{v}}\right\rangle_{1,4}\right] = 0, \tag{2.6}$$

$$\frac{\partial f^{2,4}}{\partial t} + \operatorname{div}_v\left[f^{2,4}\left\langle\dot{\vec{v}}\right\rangle_{2,4}\right] + \operatorname{div}_{\ddot{v}}\left[f^{2,4}\left\langle\dddot{\vec{v}}\right\rangle_{2,4}\right] = 0, \tag{2.7}$$

$$\frac{\partial f^{3,4}}{\partial t} + \operatorname{div}_{\dot{v}}\left[f^{3,4}\ddot{\vec{v}}\right] + \operatorname{div}_{\ddot{v}}\left[f^{3,4}\left\langle\dddot{\vec{v}}\right\rangle_{3,4}\right] = 0, \tag{2.8}$$

$$\frac{\partial f^{1,2,4}}{\partial t} + \text{div}_r\left[f^{1,2,4}\vec{v}\right] + \text{div}_v\left[f^{1,2,4}\langle\dot{\vec{v}}\rangle_{1,2,4}\right] + \text{div}_{\ddot{v}}\left[f^{1,2,4}\langle\dddot{\vec{v}}\rangle_{1,2,4}\right] = 0, \qquad (2.9)$$

$$\frac{\partial f^{2,3,4}}{\partial t} + \text{div}_v\left[f^{2,3,4}\dot{\vec{v}}\right] + \text{div}_{\dot{v}}\left[f^{2,3,4}\ddot{\vec{v}}\right] + \text{div}_{\ddot{v}}\left[f^{2,3,4}\langle\dddot{\vec{v}}\rangle_{2,3,4}\right] = 0, \qquad (2.10)$$

$$\frac{\partial f^{1,3,4}}{\partial t} + \text{div}_r\left[f^{1,3,4}\langle\vec{v}\rangle_{1,3,4}\right] + \text{div}_{\dot{v}}\left[f^{1,3,4}\ddot{\vec{v}}\right] + \text{div}_{\ddot{v}}\left[f^{1,3,4}\langle\dddot{\vec{v}}\rangle_{1,3,4}\right] = 0. \qquad (2.11)$$

Let us note the common pattern of the resulting equations. The set of equations can be conditionally divided into two types. Both types contain new equations. The first type contains equations similar to the equations in chain (i.13), in which the kinematical variables are rearranged. The first type includes equations (2.5), (2.2), (2.1), which are similar to the first Vlasov equation in chain (i.13). Equations (2.8), (2.11), and (2.3) are similar to the second equation in chain (i.13). The second type contains equations that are not similar to any of the equations in chain (i.13), for instance, equations (2.6), (2.7) and (2.4) or equations (2.9), (2.11).

Equations (2.9), (2.11) of the second type for distribution functions of the third rank transform into equations (2.6), (2.7) of the same type for functions of the second rank when they are integrated over the space of velocities and coordinates, respectively.

For clarity, we write the extensive $F^{2,\{1,2,3,4,\ldots\}}$ of the second rank, in which we select functions that satisfy different types of equations

$$F^{2,\{1,2,3,4,\ldots\}} = \begin{pmatrix} \times & f^{1,2} & f^{1,3} & f^{1,4} & f^{1,5} & \ldots \\ f^{1,2} & \times & f^{2,3} & f^{2,4} & f^{2,5} & \ldots \\ f^{1,3} & f^{2,3} & \times & f^{3,4} & f^{3,5} & \ldots \\ f^{1,4} & f^{2,4} & f^{3,4} & \times & f^{4,5} & \ldots \\ f^{1,5} & f^{2,5} & f^{3,5} & f^{4,5} & \times & \ldots \\ \ldots & \ldots & \ldots & \ldots & \ldots & \ldots \end{pmatrix}. \qquad (2.12)$$

The functions satisfying equations of the first type are green in representation (2.12), and red are ones satisfying equations of the second type. It is seen that for the extensive $F^{2,\{1,2,3,4,\ldots\}}$ of the second rank, the number of equations of the second type with increasing cardinality of set $|S|$ will be an order of magnitude larger than the number of equations of the first type. The prevalence of the number of second type equations over the number of first type equations will also be preserved for extensive of higher ranks. Note that the presence of the first type equations is an inherited feature of the hierarchical structure of the Vlasov equations chain (i.13). Indeed, the mean kinematical values in them are functions of the kinematical values of the lowest order: $\langle\dot{\vec{v}}\rangle_{1,2}$, $\langle\ddot{\vec{v}}\rangle_{2,3}$, $\langle\dddot{\vec{v}}\rangle_{3,4}$,… The second type equations contain mean kinematical values, which depend not only on the their lowest orders, but also on their higher orders: $\langle\vec{v}\rangle_{1,3}$, $\langle\vec{v}\rangle_{1,4}$, $\langle\dot{\vec{v}}\rangle_{2,4}$, $\langle\dot{\vec{v}}\rangle_{1,2,4}$,…

Summarizing our results, we obtain the dispersion chain of the Vlasov equations. Unlike chain (i.13), the dispersion chain will have two new differences. The first difference is that each equation in chain (i.13) will decompose into an infinite set of the first type equations obtained by rearranging the kinematical values of different orders. The second difference is the appearance of an infinite set of equations of the second type, containing additional terms with mean kinematical values. The dispersion chain of equations may be ranked:

<div align="center">**1st rank**</div> (2.13)

$$\frac{\partial}{\partial t}f^1 + \text{div}_r\left[f^1\langle\vec{v}\rangle_1\right] = 0, \quad \frac{\partial}{\partial t}f^2 + \text{div}_v\left[f^2\langle\dot{\vec{v}}\rangle_2\right] = 0, \quad \frac{\partial}{\partial t}f^3 + \text{div}_{\dot{v}}\left[f^3\langle\ddot{\vec{v}}\rangle_3\right] = 0,\ldots$$

<div align="center">**2nd rank**</div> (2.14)

$$\frac{\partial f^{1,2}}{\partial t} + \text{div}_r\left[f^{1,2}\vec{v}\right] + \text{div}_v\left[f^{1,2}\langle\dot{\vec{v}}\rangle_{1,2}\right] = 0,$$

$$\frac{\partial f^{1,3}}{\partial t} + \text{div}_r\left[f^{1,3}\langle\vec{v}\rangle_{1,3}\right] + \text{div}_{\dot{v}}\left[f^{1,3}\langle\ddot{\vec{v}}\rangle_{1,3}\right] = 0.$$

$$\frac{\partial f^{1,4}}{\partial t} + \text{div}_r\left[f^{1,4}\langle\vec{v}\rangle_{1,4}\right] + \text{div}_{\ddot{v}}\left[f^{1,4}\langle\dddot{\vec{v}}\rangle_{1,4}\right] = 0,$$

...

$$\frac{\partial f^{2,3}}{\partial t} + \text{div}_v\left[f^{2,3}\dot{\vec{v}}\right] + \text{div}_{\dot{v}}\left[f^{2,3}\langle\ddot{\vec{v}}\rangle_{2,3}\right] = 0,$$

$$\frac{\partial f^{2,4}}{\partial t} + \text{div}_v\left[f^{2,4}\langle\dot{\vec{v}}\rangle_{2,4}\right] + \text{div}_{\ddot{v}}\left[f^{2,4}\langle\dddot{\vec{v}}\rangle_{2,4}\right] = 0,$$

...

$$\frac{\partial f^{3,4}}{\partial t} + \text{div}_{\dot{v}}\left[f^{3,4}\ddot{\vec{v}}\right] + \text{div}_{\ddot{v}}\left[f^{3,4}\langle\dddot{\vec{v}}\rangle_{3,4}\right] = 0,$$

...

<div align="center">**3rd rank**</div> (2.15)

$$\frac{\partial f^{1,2,3}}{\partial t} + \text{div}_r\left[f^{1,2,3}\vec{v}\right] + \text{div}_v\left[f^{1,2,3}\dot{\vec{v}}\right] + \text{div}_{\dot{v}}\left[f^{1,2,3}\langle\ddot{\vec{v}}\rangle_{1,2,3}\right] = 0,$$

$$\frac{\partial f^{1,2,4}}{\partial t} + \text{div}_r\left[f^{1,2,4}\vec{v}\right] + \text{div}_v\left[f^{1,2,4}\langle\dot{\vec{v}}\rangle_{1,2,4}\right] + \text{div}_{\ddot{v}}\left[f^{1,2,4}\langle\dddot{\vec{v}}\rangle_{1,2,4}\right] = 0,$$

...

$$\frac{\partial f^{1,3,4}}{\partial t} + \text{div}_r\left[f^{1,3,4}\langle\vec{v}\rangle_{1,3,4}\right] + \text{div}_{\dot{v}}\left[f^{1,3,4}\ddot{\vec{v}}\right] + \text{div}_{\ddot{v}}\left[f^{1,3,4}\langle\dddot{\vec{v}}\rangle_{1,3,4}\right] = 0,$$

...

$$\frac{\partial f^{2,3,4}}{\partial t} + \text{div}_v\left[f^{2,3,4}\dot{\vec{v}}\right] + \text{div}_{\dot{v}}\left[f^{2,3,4}\ddot{\vec{v}}\right] + \text{div}_{\ddot{v}}\left[f^{2,3,4}\langle\dddot{\vec{v}}\rangle_{2,3,4}\right] = 0,$$

...

**Remark**

Elements $\langle\vec{\xi}^\ell\rangle$ of extensive $\vec{\Xi}^\ell_{R,S}$ may be divided into three groups. The first group is elements $\langle\vec{\xi}^\ell\rangle$ that are present in the dispersion chain and depend on the lowest order kinematical values. The second group is elements $\langle\vec{\xi}^\ell\rangle$ that are present in the dispersion chain and depend on the kinematical values of the higher and lowest orders. The third group is represented by elements $\langle\vec{\xi}^\ell\rangle$ that are not included in the dispersion chain and depend on the kinematical values of the higher and lowest order. As an example, we consider the extensive $\vec{\Xi}^4_{2,\{1,2,3,5,6\}}$:

$$\Xi^4_{2,\{1,2,3,5,6\}} = \begin{pmatrix} \times & \langle \vec{\ddot{v}} \rangle_{1,2} & \langle \vec{\ddot{v}} \rangle_{1,3} & \times & \langle \vec{\ddot{v}} \rangle_{1,5} & \langle \vec{\ddot{v}} \rangle_{1,6} \\ \langle \vec{\ddot{v}} \rangle_{1,2} & \times & \langle \vec{\ddot{v}} \rangle_{2,3} & \times & \langle \vec{\ddot{v}} \rangle_{2,5} & \langle \vec{\ddot{v}} \rangle_{2,6} \\ \langle \vec{\ddot{v}} \rangle_{1,3} & \langle \vec{\ddot{v}} \rangle_{2,3} & \times & \times & \langle \vec{\ddot{v}} \rangle_{3,5} & \langle \vec{\ddot{v}} \rangle_{3,6} \\ \times & \times & \times & \times & \times & \times \\ \langle \vec{\ddot{v}} \rangle_{1,5} & \langle \vec{\ddot{v}} \rangle_{2,5} & \langle \vec{\ddot{v}} \rangle_{3,5} & \times & \times & \langle \vec{\ddot{v}} \rangle_{5,6} \\ \langle \vec{\ddot{v}} \rangle_{1,6} & \langle \vec{\ddot{v}} \rangle_{2,6} & \langle \vec{\ddot{v}} \rangle_{3,6} & \times & \langle \vec{\ddot{v}} \rangle_{5,6} & \times \end{pmatrix}.$$

The mean kinematical values belonging to the first, second and third groups, respectively, are shown in green, red and black. The elements of the third group may be found by averaging the kinematical values of the higher ranks.

The equations of dispersion chain (2.13) - (2.15) may be rewritten in terms of differential operators $\Pi$ that allow one to single out sources of dissipation $Q$ [6].

**Definition 3.** *For rank $R$, we consider set $n = \{n_1,..,n_R\} \subset \mathbb{N}$, $n_1 \neq ... \neq n_R$. Let us define the functions:*

$$S^{n_1,...,n_R} \stackrel{det}{=} \mathrm{Ln}\, f^{n_1,...,n_R}, \qquad Q^p_{n_1,...,n_R} \stackrel{det}{=} \mathrm{div}_{\xi^p} \langle \vec{\xi}^{p+1} \rangle_{n_1,...,n_R}, \qquad (2.16)$$

*where $p \in n$, and functions $f^{n_1,...,n_R}$ have no zero values. Let us call function $Q^p_{n_1,...,n_R}$ the density of dissipation sources of mean kinematical value $\langle \vec{\xi}^{p+1} \rangle_{n_1,...,n_R}$.*

**Definition 4.** *Let us define the extensive of operators $\Pi^R = \{\pi_{n_1,...,n_R}\}$ of rank $R$, where $\pi_{n_1,...,n_R} \stackrel{det}{=} \dfrac{d_{n_1,...,n_R}}{dt}$:*

at $R = 0$: $\quad \pi_0 = \dfrac{d}{dt},$

at $R = 1$: $\quad \pi_n = \dfrac{\partial}{\partial t} + \langle \vec{\xi}^{n+1} \rangle_n \nabla_{\xi^n},$

at $R = 2$ $\quad \pi_{n,n+1} = \dfrac{\partial}{\partial t} + \vec{\xi}^{n+1}\nabla_{\xi^n} + \langle \vec{\xi}^{n+2} \rangle_{n,n+1} \nabla_{\xi^{n+1}},$

$\quad \pi_{n,n+k} = \dfrac{\partial}{\partial t} + \langle \vec{\xi}^{n+1} \rangle_{n,n+k} \nabla_{\xi^n} + \langle \vec{\xi}^{n+1+k} \rangle_{n,n+k} \nabla_{\xi^{n+k}},$ (2.17)

at $R = 3$

$\quad \pi_{n,n+1,n+2} = \dfrac{\partial}{\partial t} + \vec{\xi}^{n+1}\nabla_{\xi^n} + \vec{\xi}^{n+2}\nabla_{\xi^{n+1}} + \langle \vec{\xi}^{n+3} \rangle_{n,n+1,n+2} \nabla_{\xi^{n+2}},$

$\quad \pi_{n,n+1,n+1+k} = \dfrac{\partial}{\partial t} + \vec{\xi}^{n+1}\nabla_{\xi^n} + \langle \vec{\xi}^{n+2} \rangle_{n,n+1,n+1+k} \nabla_{\xi^{n+1}} + \langle \vec{\xi}^{n+2+k} \rangle_{n,n+1,n+1+k} \nabla_{\xi^{n+1+k}},$

$\quad \pi_{n,n+s,n+s+1} = \dfrac{\partial}{\partial t} + \langle \vec{\xi}^{n+1} \rangle_{n,n+s,n+s+1} \nabla_{\xi^n} + \vec{\xi}^{n+s+1}\nabla_{\xi^{n+s}} + \langle \vec{\xi}^{n+s+2} \rangle_{n,n+s,n+s+1} \nabla_{\xi^{n+s+1}},$

$$\pi_{n,n+s,n+s+k} = \frac{\partial}{\partial t} + \left\langle \vec{\xi}^{n+1} \right\rangle_{n,n+s,n+s+k} \nabla_{\xi^n} + \left\langle \vec{\xi}^{n+1+s} \right\rangle_{n,n+s,n+s+k} \nabla_{\xi^{n+s}} + \left\langle \vec{\xi}^{n+1+s+k} \right\rangle_{n,n+s,n+s+k} \nabla_{\xi^{n+s+k}},$$

....

where $k, s \in \{2, 3, ...\}$.

Using definitions (2.16) and (2.17) and taking into account the independence of kinematical values $\vec{r}, \vec{v}, \dot{\vec{v}}, \ddot{\vec{v}}, ...$, dispersion chain (2.13)-(2.15) will be of the form:

**1st rank** (2.18)
$$\pi_n S^n = -Q_n^n.$$

**2nd rank** (2.19)
$$\pi_{n,n+1} S^{n,n+1} = -Q_{n,n+1}^{n+1},$$
$$\pi_{n,n+k} S^{n,n+k} = -\left(Q_{n,n+k}^n + Q_{n,n+k}^{n+k}\right).$$

**3rd rank** (2.20)
$$\pi_{n,n+1,n+2} S^{n,n+1,n+2} = -Q_{n,n+1,n+2}^{n+2},$$
$$\pi_{n,n+1,n+1+k} S^{n,n+1,n+1+k} = -\left(Q_{n,n+1,n+1+k}^{n+1} + Q_{n,n+1,n+1+k}^{n+1+k}\right),$$
$$\pi_{n,n+s,n+s+1} S^{n,n+s,n+s+1} = -\left(Q_{n,n+s,n+s+1}^n + Q_{n,n+s,n+s+1}^{n+s+1}\right),$$
$$\pi_{n,n+s,n+s+k} S^{n,n+s,n+s+k} = -\left(Q_{n,n+s,n+s+k}^n + Q_{n,n+s,n+s+k}^{n+s} + Q_{n,n+s,n+s+k}^{n+s+k}\right).$$

...

where $k, s \in \{2, 3, ...\}$.

Note that, as the rank of extensive $\Pi^R$ increases, the number of mean kinematical values $\left\langle \vec{\xi}^n \right\rangle$ in operators $\pi_{n_1,...,n_R}$ increases. Operators $\pi_{n,n+1,n+2,...,n+R-1}$ correspond to the first group of equations for distribution functions $f_{n,n+1,...,n+R-1}$. Let us consider extensive $\Pi^2$:

$$\Pi^2 = \begin{pmatrix} \times & \pi_{1,2} & \pi_{1,3} & \pi_{1,4} & \pi_{1,5} & ... \\ \pi_{1,2} & \times & \pi_{2,3} & \pi_{2,4} & \pi_{2,5} & ... \\ \pi_{1,3} & \pi_{2,3} & \times & \pi_{3,4} & \pi_{3,5} & ... \\ \pi_{1,4} & \pi_{2,4} & \pi_{3,4} & \times & \pi_{4,5} & ... \\ \pi_{1,5} & \pi_{2,5} & \pi_{3,5} & \pi_{4,5} & \times & ... \\ ... & ... & ... & ... & ... & ... \end{pmatrix}. \quad (2.21)$$

The operators corresponding to the first group of equations are marked in green in expression (2.21), and the ones corresponding to the second group – in red. Note that the extensive $F^2$ (2.12) has a similar distribution of elements.

The dispersion chain, written in the form (2.18) - (2.20), clearly shows that the presence of sources of dissipation $Q_{n_1,...,n_R}^p \neq 0$ leads to non-conservation of probability density $f^{n_1,...,n_R}$ along the phase trajectories. Indeed, operators $\pi_{n_1,...,n_R}$ actually set a «total derivative» in time from distribution function $f^{n_1,...,n_R}$, which is equal to $Q_{n_1,...,n_R}^p$.

From expressions (2.18) - (2.20) it is seen that the equations of the first group for functions $S^n$, $S^{n,n+1}$, $S^{n,n+1,n+2}$,... have sources of dissipation only in one mean kinematical value. In this case, sources of dissipations $Q_n^n$, $Q_{n,n+1}^{n+1}$, $Q_{n,n+1,n+2}^{n+2}$,.... are functions of the kinematical values of the lowest orders. Equations from the second group $S^{n,n+k}$, $S^{n,n+1,n+1+k}$, $S^{n,n+s,n+s+1}$, $S^{n,n+s,n+s+k}$,... depending on the rank have two, three and more sources of dissipations. Sources $Q_{n,n+k}^n$, $Q_{n,n+k}^{n+k}$, $Q_{n,n+1,n+1+k}^{n+1}$, $Q_{n,n+1,n+1+k}^{n+1+k}$, $Q_{n,n+s,n+s+k}^n$, $Q_{n,n+s,n+s+k}^{n+s}$, $Q_{n,n+s,n+s+k}^{n+s+k}$, … depend not only on the lowest kinematical values, but on higher ones.

**Remark**

As can be seen from dispersion chain (2.18) - (2.20), distribution functions of the same rank may be determined by various sets of dissipation sources. For instance, the second rank functions $f^{n,n+1}$ are determined by one dissipation source $Q_{n,n+1}^{n+1}$ (2.19), and the second rank functions $f^{n,n+k}$, $k > 1$ are determined by two dissipation sources $Q_{n,n+k}^n$ and $Q_{n,n+k}^{n+k}$. The kinematical values, which functions $f^{n,n+1}$ and $f^{n,n+k}$ depend on, are independent, then why does such an «inhomogeneity» present?

In initial chain of the Vlasov equations (i.13), the transition from the equation for the distribution function with rank $R+1$ to the equation for the distribution function with rank $R$ occurs by integration over the corresponding phase subspace. With such a transition, information on the mean kinematical values is lost. This issue is considered in Introduction (see (i.1), (i.6), (i.11)) using the example of the first three equations from the Vlasov chain (i.13). Indeed, such a picture is observed for distribution functions from the first group, that is, for $f^{n,n+1}$. Sources of dissipation $Q_{n,n+1}^{n+1}$ depend only on the kinematical values of the lowest orders and there is only one such source. Despite the fact that kinematical values $\vec{r}, \vec{v}, \dot{\vec{v}}, \ddot{\vec{v}},...$ are independent, they have a strict hierarchy. Dispersion chain (2.18) - (2.20) contains functions $f^{n,n+k}$ belonging to the second group in which the kinematical values violate the hierarchy. The violation of the hierarchy leads to the influence of dissipation sources ($Q_{n,n+k}^n$, $Q_{n,n+k}^{n+k}$) from different phase subspaces on distribution function $f^{n,n+k}$. In accordance with equations (2.19), the extensive of the second rank of the sources of dissipation may be represented as follows:

$$\begin{pmatrix} \times & Q_{1,2}^2 & Q_{1,3}^1 + Q_{1,3}^3 & Q_{1,4}^1 + Q_{1,4}^4 & Q_{1,5}^1 + Q_{1,5}^5 \\ Q_{1,2}^2 & \times & Q_{2,3}^3 & Q_{2,4}^2 + Q_{2,4}^4 & Q_{2,5}^2 + Q_{2,5}^5 \\ Q_{1,3}^1 + Q_{1,3}^3 & Q_{2,3}^3 & \times & Q_{3,4}^4 & Q_{3,5}^3 + Q_{3,5}^5 \\ Q_{1,4}^1 + Q_{1,4}^4 & Q_{2,4}^2 + Q_{2,4}^4 & Q_{3,4}^4 & \times & Q_{4,5}^5 \\ Q_{1,5}^1 + Q_{1,5}^5 & Q_{2,5}^2 + Q_{2,5}^5 & Q_{3,5}^3 + Q_{3,5}^5 & Q_{4,5}^5 & \times \end{pmatrix} \quad (2.22)$$

The single sources of dissipation included in the equations from the first group are shown in green. The red color corresponds to the mixed distribution functions from the second group.

**§3 Conservation laws**

Let us show that from dispersion chain (2.13) - (2.15) it is possible to obtain an infinite chain of conservation laws for «matter», «momentum» and «energy». The quotation marks indicate formal similarities with the known concepts. For the coordinate space, the dispersion chain gives the known conservation laws [1]. When considering higher kinematical values, the mathematical form of the equations is also preserved.

Let us consider first rank equations (2.13) for distribution functions $f^n$. The first equation in (2.13) coincides with the continuity equation for mass/charge density $f^1(\vec{r},t)$. The rest of the equations from (2.13) have a mathematically similar form and are written for densities $f^2(\vec{v},t)$, $f^3(\dot{\vec{v}},t)$,.... As noted above, first rank equations (2.13) are obtained from second rank equations (2.14) by integration over the corresponding phase subspaces. Equations (2.13) can formally be interpreted as equations expressing the law of conservation of «matter».

Equations of the second rank (2.14) consist of two groups $f^{n,n+1}$ (first group) and $f^{n,n+k}$, $k>1$ (second group). Let us first consider the equations of the first group, which may be written in the form:

$$\frac{\partial f^{n,n+1}}{\partial t} + \mathrm{div}_{\xi^n}\left[f^{n,n+1}\vec{\xi}^{n+1}\right] + \mathrm{div}_{\xi^{n+1}}\left[f^{n,n+1}\left\langle\vec{\xi}^{n+2}\right\rangle_{n,n+1}\right] = 0. \qquad (3.1)$$

**Theorem 1.** *Let distribution function of the second rank $f^{n,n+1}$ from the first group satisfies equation (3.1), then the conservation laws are valid for mean kinematical value $\left\langle\vec{\xi}^{n+1}\right\rangle_n$:*

$$\pi_n\left\langle\xi_\alpha^{n+1}\right\rangle_n = \left[\frac{\partial}{\partial t} + \left\langle\xi_\beta^{n+1}\right\rangle_n\frac{\partial}{\partial\xi_\beta^n}\right]\left\langle\xi_\alpha^{n+1}\right\rangle_n = -\frac{1}{f^n}\frac{\partial P_{\alpha\beta}^{n+1}}{\partial\xi_\beta^n} + \left\langle\xi_\alpha^{n+2}\right\rangle_n, \qquad (3.2)$$

$$\frac{\partial}{\partial t}\left[\frac{f^n}{2}\left\langle\xi^{n+1}\right\rangle_n^2 + \frac{1}{2}\mathrm{Tr}\,P_{\alpha\alpha}^{n+1}\right] + \frac{\partial}{\partial\xi_\beta^n}\left[\frac{f^n}{2}\left\langle\xi^{n+1}\right\rangle_n^2\left\langle\xi_\beta^{n+1}\right\rangle_n + \frac{1}{2}\left\langle\xi_\beta^{n+1}\right\rangle_n\mathrm{Tr}\,P_{\alpha\alpha}^{n+1} + \left\langle\xi_\alpha^{n+1}\right\rangle_n P_{\alpha\beta}^{n+1} + \frac{1}{2}\mathrm{Tr}\,P_{\alpha\alpha\beta}^{n+1}\right] =$$
$$= \int_{\Omega_{n+1}} f^{n,n+1}\left\langle\xi_\alpha^{n+2}\right\rangle_{n,n+1}\xi_\alpha^{n+1}\,d^3\xi^{n+1}, \qquad (3.3)$$

*where $P_{\alpha\beta}^{n+1}$ and $P_{\alpha\beta\mu}^{n+1}$ are momenta of the second and third order, respectively, for kinematical value $\vec{\xi}^{n+1}$:*

$$P_{\alpha\beta}^{n+1} \stackrel{\mathrm{det}}{=} \int_{\Omega_{n+1}}\left(\xi_\alpha^{n+1} - \left\langle\xi_\alpha^{n+1}\right\rangle_n\right)\left(\xi_\beta^{n+1} - \left\langle\xi_\beta^{n+1}\right\rangle_n\right)f^{n,n+1}d^3\xi^{n+1}, \qquad (3.4)$$

$$P_{\alpha\beta\mu}^{n+1} \stackrel{\mathrm{det}}{=} \int_{\Omega_{n+1}}\left(\xi_\alpha^{n+1} - \left\langle\xi_\alpha^{n+1}\right\rangle_n\right)\left(\xi_\beta^{n+1} - \left\langle\xi_\beta^{n+1}\right\rangle_n\right)\left(\xi_\mu^{n+1} - \left\langle\xi_\mu^{n+1}\right\rangle_n\right)f^{n,n+1}d^3\xi^{n+1}. \qquad (3.5)$$

The proof of Theorem 1 is given in Appendix A.

Expression (3.2) at $n=1$ transforms into known equation of hydrodynamics (i.7), which can be formally called the law of conservation of «momentum». Second equation (3.3) at $n=1$ coincides with the energy conservation law in hydrodynamics [1]:

$$\frac{\partial}{\partial t}\left[\frac{f^1}{2}|\langle\vec{v}\rangle|^2 + \frac{1}{2}\mathrm{Tr}\,P_{\alpha\alpha}^2\right] + \frac{\partial}{\partial x_\beta}\left[\frac{f^1}{2}|\langle\vec{v}\rangle|^2\langle v_\beta\rangle + \frac{1}{2}\langle v_\beta\rangle\mathrm{Tr}\,P_{\alpha\alpha}^2 + \langle v_\alpha\rangle P_{\alpha\beta}^2 + \frac{1}{2}\mathrm{Tr}\,P_{\alpha\alpha\beta}^2\right] = \int_{(\infty)}f^{1,2}\langle\dot{v}_\alpha\rangle v_\alpha\,d^3v, \qquad (3.6)$$

where summand $\frac{f^1}{2}|\langle\vec{v}\rangle|^2$ determines the density of kinetic energy; $\frac{1}{2}\text{Tr}\,P^2_{\alpha\alpha}$ corresponds to the density of the internal energy; $\frac{f^1}{2}|\langle\vec{v}\rangle|^2\langle v_\beta\rangle$ sets the flow of the kinetic energy; $\frac{1}{2}\langle v_\beta\rangle\text{Tr}\,P^2_{\alpha\alpha}$ – the flow of the internal energy; $\langle v_\alpha\rangle P^2_{\alpha\beta}$ characterizes the gravitation work; $\frac{1}{2}\text{Tr}\,P^2_{\alpha\alpha\beta}$ – the heat flow. The right-hand side of equation (3.6) $m\int_{(\infty)} f^{1,2}\langle\dot{v}_\alpha\rangle v_\alpha d^3v$ is the mean value of the external forces. Thus, equation (3.3) may be formally called the «energy» conservation law.

Let us consider the second group of equations for the distribution functions $f^{n,n+k}$, $k>1$ of the second rank. According to (2.14) and (2.19) such equations are of the form:

$$\frac{\partial f^{n,n+k}}{\partial t}+\text{div}_{\xi^n}\left[f^{n,n+k}\langle\vec{\xi}^{n+1}\rangle_{n,n+k}\right]+\text{div}_{\xi^{n+k}}\left[f^{n,n+k}\langle\vec{\xi}^{n+k+1}\rangle_{n,n+k}\right]=0. \tag{3.7}$$

**Theorem 2.** *Let the distribution function $f^{n,n+k}$, $k>1$ of the second rank from the second group, satisfy equation (3.7), then the conservation laws are valid for the mean kinematical value $\langle\vec{\xi}^{n+k}\rangle_n$:*

$$\pi_n\langle\xi_\alpha^{n+k}\rangle_n=\left[\frac{\partial}{\partial t}+\langle\xi_\beta^{n+1}\rangle_n\frac{\partial}{\partial\xi_\beta^n}\right]\langle\xi_\alpha^{n+k}\rangle_n=-\frac{1}{f^n}\frac{\partial P_{\beta\alpha}^{n+1,n+k}}{\partial\xi_\beta^n}+\langle\xi_\alpha^{n+k+1}\rangle_n, \tag{3.8}$$

$$\frac{\partial}{\partial t}\left[\frac{f^n}{2}\langle\xi^{n+k}\rangle_n^2+\frac{1}{2}\text{Tr}\,P_{\alpha\alpha}^{n+k}\right]+$$

$$+\frac{\partial}{\partial\xi_\beta^n}\left[\frac{f^n}{2}\langle\xi^{n+k}\rangle_n^2\langle\xi_\beta^{n+1}\rangle_n+\frac{1}{2}\langle\xi_\beta^{n+1}\rangle_n\text{Tr}\,P_{\alpha\alpha}^{n+k}+P_{\beta\alpha}^{n+1,n+k}\langle\xi_\alpha^{n+k}\rangle_n+\frac{1}{2}\text{Tr}\,P_{\beta\alpha\alpha}^{n+1,n+k,n+k}\right]= \tag{3.9}$$

$$=\int_{\Omega_{n+k}} f^{n,n+k}\langle\xi_\alpha^{n+k+1}\rangle_{n,n+k}\xi_\alpha^{n+k}d^3\xi^{n+k},$$

where $P_{\alpha\beta}^{n+1,n+k}$ and $P_{\beta\alpha\alpha}^{n+1,n+k,n+k}$ are momenta of the second and third orders, respectively, for kinematical values $(\vec{\xi}^{n+1},\vec{\xi}^{n+k})$:

$$P_{\alpha\beta}^{n+1,n+k}\stackrel{\text{det}}{=}\int_{\Omega_{n+1}}\int_{\Omega_{n+k}}\left(\xi_\alpha^{n+1}-\langle\xi_\alpha^{n+1}\rangle_n\right)\left(\xi_\beta^{n+k}-\langle\xi_\beta^{n+k}\rangle_n\right)f^{n,n+1,n+k}d^3\xi^{n+1}d^3\xi^{n+k}, \tag{3.10}$$

$$P_{\beta\alpha\alpha}^{n+1,n+k,n+k}\stackrel{\text{det}}{=}\int_{\Omega_{n+1}}\int_{\Omega_{n+k}}\left(\xi_\beta^{n+1}-\langle\xi_\beta^{n+1}\rangle_n\right)\left(\xi_\alpha^{n+k}-\langle\xi_\alpha^{n+k}\rangle_n\right)^2 f^{n,n+1,n+k}d^3\xi^{n+1}d^3\xi^{n+k}. \tag{3.11}$$

The proof of Theorem 2 is given in Appendix A.

Equation (3.9) expresses the law of conservation of «energy» for the second group of kinematic quantities. Comparison of expressions (3.3) and (3.9) shows that the law of conservation of «energy» for the first and second groups has a similar form. At $k=1$, equation (3.9) formally transforms into equation (3.3). Equations (3.8) and (3.9) contain values $P_{\beta\alpha}^{n+1,n+k}$ and $P_{\beta\alpha\alpha}^{n+1,n+k,n+k}$ that characterize their belonging to the second group. In contrast to the momenta $P_{\alpha\beta}^{n+1}$ and $P_{\alpha\alpha\beta}^{n+1}$, included in equations (3.2) and (3.3) of the first group, the momenta $P_{\beta\alpha}^{n+1,n+k}$ and $P_{\beta\alpha\alpha}^{n+1,n+k,n+k}$ are written for kinematical values of different orders (the second group). Resulting equation (3.8) and equation (3.2) have a similar form. At $k=1$, tensor $P_{\beta\alpha}^{n+1,n+k}$ degenerates into tensor $P_{\alpha\beta}^{n+1}$ and equation (3.8) (for the second group) transforms into equation (3.2) (for the first group).

For the distribution functions of the third and highest ranks, one can make similar calculations and obtain equations of «conservation» laws of the type (3.2) / (3.8) and (3.3) / (3.9). The distribution functions $f^{n_1,n_2}$ of the second rank considered above depended on two kinematical values $\vec{\xi}^{n_1}$ and $\vec{\xi}^{n_2}$, therefore, the integration in expressions (A.1), (A.5), (A.11), (A.16) was performed only over one variable. As a result, conservation laws (3.2), (3.3), (3.8), (3.9) were obtained for functions of one kinematic variable.

Distribution functions $f^{n_1,n_2,n_3}$ of the third rank depend on three kinematical values $\vec{\xi}^{n_1}$, $\vec{\xi}^{n_2}$, $\vec{\xi}^{n_3}$. Therefore, integration can be performed over one or two kinematical values. When integrating over two variables, conservation laws are obtained that depend on one kinematical value. Such conservation laws coincide with obtained equations (3.2), (3.3), (3.8) and (3.9). New equations of conservation laws can be obtained by integrating equations of the third rank over only one kinematic variable.

Let us consider the distribution functions $f^{n,n+1,n+2}$ of the third rank from the first group (one source of dissipation $Q_{n,n+1,n+2}^{n+2}$) (2.15), (2.20).

$$\frac{\partial f^{n,n+1,n+2}}{\partial t} + \xi_\beta^{n+1} \frac{\partial f^{n,n+1,n+2}}{\partial \xi_\beta^n} + \xi_\beta^{n+2} \frac{\partial f^{n,n+1,n+2}}{\partial \xi_\beta^{n+1}} + \frac{\partial}{\partial \xi_\beta^{n+2}} \left[ f^{n,n+1,n+2} \left\langle \xi_\beta^{n+3} \right\rangle_{n,n+1,n+2} \right] = 0. \quad (3.12)$$

**Theorem 3.** *Let the distribution function $f^{n,n+1,n+2}$ of the third rank belonging to the first group satisfy equation (3.12), then for mean kinematical values $\left\langle \vec{\xi}^n \right\rangle_{n+1,n+2}$, $\left\langle \vec{\xi}^{n+1} \right\rangle_{n,n+2}$ and $\left\langle \vec{\xi}^{n+2} \right\rangle_{n,n+1}$ the following equations of motion are valid:*

$$\pi_{n+1,n+2} \left\langle \xi_\alpha^n \right\rangle_{n+1,n+2} = -\frac{1}{f^{n+1,n+2}} \frac{\partial P_{\alpha\beta}^{n,n+3}}{\partial \xi_\beta^{n+2}} + \xi_\alpha^{n+1}, \quad (3.13)$$

$$\pi_{n,n+2} \left\langle \xi_\alpha^{n+1} \right\rangle_{n,n+2} = -\frac{1}{f^{n,n+2}} \left( \frac{\partial P_{\alpha\beta}^{n+1}}{\partial \xi_\beta^n} + \frac{\partial P_{\alpha\beta}^{n+1,n+3}}{\partial \xi_\beta^{n+2}} \right) + \xi_\alpha^{n+2}, \quad (3.14)$$

$$\pi_{n,n+1} \left\langle \xi_\alpha^{n+2} \right\rangle_{n,n+1} = -\frac{1}{f^{n,n+1}} \frac{\partial P_{\alpha\beta}^{n+2}}{\partial \xi_\beta^{n+1}} + \left\langle \xi_\alpha^{n+3} \right\rangle_{n,n+1}, \quad (3.15)$$

where $P_{\alpha\beta}^{n+2}$, $P_{\alpha\beta}^{n,n+3}$, $P_{\alpha\beta}^{n+1}$, $P_{\alpha\beta}^{n+1,n+3}$ are momenta of the second order defined as $P_{\alpha\beta}^{n+2} = P_{\alpha\beta}^{n+2} \left( \vec{\xi}^n, \vec{\xi}^{n+1} \right) \overset{\text{det}}{=} P_{\alpha\beta}^{n+2} (n, n+1)$. Similarly, $P_{\alpha\beta}^{n,n+3} (n+1, n+2)$, $P_{\alpha\beta}^{n+1} (n, n+2)$ and $P_{\alpha\beta}^{n+1,n+3} (n, n+2)$.

The proof of Theorem 3 is given in Appendix A.

Equations of motion (3.13) and (3.15) depend on the distribution functions of the second rank from the first group. Functions $f^{n,n+1}$ and $f^{n+1,n+2}$ satisfy dispersion equations (2.19):

$$\pi_{n,n+1} S^{n,n+1} = -Q^{n+1}_{n,n+1}, \qquad \pi_{n+1,n+2} S^{n+1,n+2} = -Q^{n+2}_{n+1,n+2},$$

which contain only one source of dissipation. Value $Q^{n+1}_{n,n+1}$ corresponds to the density of the field sources $\langle \vec{\xi}^{n+2} \rangle_{n,n+1}$ which tensor $P^{n+2}_{\alpha\beta}$ (3.15) depends on. Similarly, value $Q^{n+2}_{n+1,n+2}$ determines the density of the sources of the field $\langle \vec{\xi}^{n+3} \rangle_{n+1,n+2}$ that enters tensor $P^{n,n+3}_{\alpha\beta}$. In contrast to equation (3.15), equation (3.13) is written for mean kinematical values $\langle \xi^n_\alpha \rangle_{n+1,n+2}$, which depends on the higher order kinematical values (second group). As a result, tensor $P^{n,n+3}_{\alpha\beta}$ in equation (3.13) depends on kinematical values of different orders.

Equation of motion (3.14) contains distribution function $f^{n,n+2}$ from the second group satisfying dispersion equation (2.19)

$$\pi_{n,n+2} S^{n,n+2} = -\left( Q^n_{n,n+2} + Q^{n+2}_{n,n+2} \right),$$

which contains two dissipation sources. Values $Q^n_{n,n+2}$ and $Q^{n+2}_{n,n+2}$ are determined by vector fields $\langle \vec{\xi}^{n+1} \rangle_{n,n+2}$ and $\langle \vec{\xi}^{n+3} \rangle_{n,n+2}$ respectively, which tensors $P^{n+1}_{\alpha\beta}$ and $P^{n+1,n+3}_{\alpha\beta}$ depend on (3.14). Equations (3.13) and (3.14) are written for kinematical values $\langle \xi^n_\alpha \rangle_{n+1,n+2}$ and $\langle \xi^{n+1}_\alpha \rangle_{n,n+2}$ from the second group but having different numbers of dissipation sources. Equations (3.15) and (3.2) are alike, since they belong to the first group with only one source of dissipations.

In the second group, there are several types of distribution functions of the third rank: $f^{n,n+1,n+1+k}$, $f^{n,n+s,n+s+1}$, $f^{n,n+s,n+s+k}$, where $k,s \in [2,+\infty)$. Without restricting the generality, we consider only the equation for distribution function $f^{n,n+1,n+1+k}$ of the third rank

$$\frac{\partial f^{n,n+1,n+1+k}}{\partial t} + \frac{\partial}{\partial \xi^n_\beta} \left[ \xi^{n+1}_\beta f^{n,n+1,n+1+k} \right] + \frac{\partial}{\partial \xi^{n+1}_\beta} \left[ f^{n,n+1,n+1+k} \langle \xi^{n+2}_\beta \rangle_{n,n+1,n+1+k} \right] +$$
$$+ \frac{\partial}{\partial \xi^{n+1+k}_\beta} \left[ f^{n,n+1,n+1+k} \langle \xi^{n+2+k}_\beta \rangle_{n,n+1,n+1+k} \right] = 0. \tag{3.16}$$

**Theorem 4.** *Let distribution function of the third rank $f^{n,n+1,n+1+k}$, $k > 1$ from the second group satisfy equation (3.16), then for mean kinematical values $\langle \vec{\xi}^n \rangle_{n+1,n+1+k}$, $\langle \vec{\xi}^{n+1} \rangle_{n,n+1+k}$ and $\langle \vec{\xi}^{n+1+k} \rangle_{n,n+1}$ the following equations of motion are valid:*

$$\pi_{n+1,n+1+k} \langle \xi^n_\alpha \rangle_{n+1,n+1+k} = -\frac{1}{f^{n+1,n+1+k}} \left( \frac{\partial P^{n,n+2}_{\alpha\beta}}{\partial \xi^{n+1}_\beta} + \frac{\partial P^{n,n+2+k}_{\alpha\beta}}{\partial \xi^{n+1+k}_\beta} \right) + \xi^{n+1}_\alpha, \tag{3.17}$$

$$\pi_{n,n+1+k} \langle \xi^{n+1}_\alpha \rangle_{n,n+1+k} = -\frac{1}{f^{n,n+1+k}} \left( \frac{\partial P^{n+1}_{\alpha\beta}}{\partial \xi^n_\beta} + \frac{\partial P^{n+1,n+2+k}_{\alpha\beta}}{\partial \xi^{n+1+k}_\beta} \right) + \langle \xi^{n+2}_\alpha \rangle_{n,n+1+k}, \tag{3.18}$$

$$\pi_{n,n+1} \left\langle \xi_\alpha^{n+1+k} \right\rangle_{n,n+1} = -\frac{1}{f^{n,n+1}} \frac{\partial P_{\beta\alpha}^{n+2,n+1+k}}{\partial \xi_\beta^{n+1}} + \left\langle \xi_\alpha^{n+2+k} \right\rangle_{n,n+1}, \qquad (3.19)$$

where $P_{\alpha\beta}^{n,n+2}$, $P_{\alpha\beta}^{n,n+2+k}$, $P_{\alpha\beta}^{n+1}$, $P_{\alpha\beta}^{n+1,n+2+k}$ and $P_{\alpha\beta}^{n+2,n+1+k}$ are momenta of the second order defined as $P_{\alpha\beta}^{n,n+2}(n+1,n+1+k)$, $P_{\alpha\beta}^{n,n+2+k}(n+1,n+1+k)$, $P_{\alpha\beta}^{n+1}(n,n+1+k)$, $P_{\alpha\beta}^{n+2,n+1+k}(n,n+1+k)$ and $P_{\alpha\beta}^{n+2,n+1+k}(n,n+1)$.

The proof of Theorem 4 is given in Appendix A.

Note that at $k=1$, equations (3.18) and (3.19) formally degenerate into equations (3.14) and (3.15), respectively. At $k=1$, equation (3.17) does not degenerate into equation (3.13), since the dispersion equation for function $f^{n+1,n+1+k}$ has two sources of dissipation: $\pi_{n+1,n+1+k} S^{n+1,n+1+k} = -\left( Q_{n+1,n+1+k}^{n+1} + Q_{n+1,n+1+k}^{n+1+k} \right)$.

The obtained equations of motion give an interesting interpretation of tensors $P_{\alpha\beta}^\lambda$. For instance, from equations of the form (3.2), (3.15) it follows that:

$$\frac{\partial}{\partial \xi_\beta^n} P_{\alpha\beta}^{n+1}(n) = f^n \left[ \left\langle \xi_\alpha^{n+2} \right\rangle_n - \pi_n \left\langle \xi_\alpha^{n+1} \right\rangle_n \right],$$

$$\frac{\partial}{\partial \xi_\beta^{n+1}} P_{\alpha\beta}^{n+2}(n,n+1) = f^{n,n+1} \left[ \left\langle \xi_\alpha^{n+3} \right\rangle_{n,n+1} - \pi_{n,n+1} \left\langle \xi_\alpha^{n+2} \right\rangle_{n,n+1} \right], \qquad (3.20)$$

$$\frac{\partial}{\partial \xi_\beta^{n+\lambda}} P_{\alpha\beta}^{n+1+\lambda}(n,...,n+\lambda) = f^{n,...,n+\lambda} \left[ \left\langle \xi_\alpha^{n+2+\lambda} \right\rangle_{n,...,n+\lambda} - \pi_{n,...,n+\lambda} \left\langle \xi_\alpha^{n+1+\lambda} \right\rangle_{n,...,n+\lambda} \right],$$

...

where $\lambda = 0,1,...$ Thus, value $\frac{\partial}{\partial \xi_\beta^{n+\lambda}} P_{\alpha\beta}^{n+1+\lambda}$ determines the difference between mean kinematical value $\left\langle \xi_\alpha^{n+2+\lambda} \right\rangle$ of order $n+2+\lambda$ and the total derivative with respect to time $\pi$ of mean kinematical value $\left\langle \xi_\alpha^{n+1+\lambda} \right\rangle$ of order $n+1+\lambda$.

**Theorem 5** *If distribution function* $f^{n,...n+\lambda,n+1+\lambda}$, $\lambda = 0,1,...$ *is even in variable* $\vec{\xi}^{n+\lambda}$, *that is*

$$f^{n,...,n+1+\lambda}\left( \vec{\xi}^n,...,-\vec{\xi}^{n+\lambda},\vec{\xi}^{n+1+\lambda} \right) = f^{n,...,n+1+\lambda}\left( \vec{\xi}^n,...,\vec{\xi}^{n+\lambda},\vec{\xi}^{n+1+\lambda} \right), \qquad (3.21)$$

*or* $P_{\alpha\beta}^{n+1+\lambda} = const$, *then*

$$\left\langle \xi_\alpha^{n+2+\lambda} \right\rangle_{n,...,n+\lambda-1} = \pi_{n,...,n+\lambda-1} \left\langle \xi_\alpha^{n+1+\lambda} \right\rangle_{n,...,n+\lambda-1}. \qquad (3.22)$$

The proof of Theorem 5 is given in Appendix A.

**Corollary 1.** *When the condition* $P_{\alpha\beta}^{n+1+\lambda} = const$, $\lambda > 0$ *of Theorem 5 is satisfied, the average value of the kinematical value may be determined as*

$$\left\langle \xi_\alpha^{n+2+\lambda} \right\rangle_{n,\ldots,n+\lambda-1} = \frac{1}{f^{n,\ldots,n+\lambda-1}} \int_{\Omega_{n+\lambda}} f^{n,\ldots,n+\lambda} \pi_{n,\ldots,n+\lambda} \left\langle \xi_\alpha^{n+1+\lambda} \right\rangle_{n,\ldots,n+\lambda} d^3\xi^{n+\lambda}. \quad (3.23)$$

**Example**

As an example illustrating the formulation of Theorem 5, let us consider the one-velocity distribution function of the second rank $f^{1,2}$ of the form

$$f^{1,2} = f_2(\vec{r},\vec{v},t) = \rho(\vec{r},t)\delta(\vec{v} - \langle\vec{v}\rangle(\vec{r},t)). \quad (3.24)$$

Function (3.24) corresponds to a system of particles with coordinate mass density $\rho(\vec{r},t) = f^1(\vec{r},t)$ and vector velocity field $\langle\vec{v}\rangle(\vec{r},t)$. We calculate tensor $P^2_{\alpha\beta}$, we obtain:

$$P^2_{\alpha\beta}(\vec{r},t) = \rho(\vec{r},t) \int_{\Omega_2} \left(v_\alpha - \langle v_\alpha \rangle_1\right)\left(v_\beta - \langle v_\beta \rangle_1\right) \delta(\vec{v} - \langle\vec{v}\rangle(\vec{r},t)) d^3v = 0. \quad (3.25)$$

Expression (3.25) satisfies the condition of Theorem 5, therefore,

$$\left\langle \dot{\vec{v}} \right\rangle_1 = \pi_1 \langle\vec{v}\rangle_1 = \left[\frac{\partial}{\partial t} + \langle\vec{v}\rangle_1 \nabla_r\right]\langle\vec{v}\rangle_1. \quad (3.26)$$

Representation (3.26) is an analogue of hydrodynamic equation (i.7) ((3.2), $n=1$) in the absence of pressure.

As another example of function $f^{1,2}$, we consider the Wigner function [9], used in describing quantum systems in a phase space. For mean kinematical value $\langle\dot{\vec{v}}\rangle_{1,2}$, one can use the Vlasov-Moyal approximation (i.9). From the second equation in (3.20), we obtain

$$\frac{1}{f^{1,2}} \frac{\partial P^3_{\alpha\beta}}{\partial v_\beta} = \left\langle \ddot{v}_\alpha \right\rangle_{1,2} - \sum_{k=0}^{+\infty} \frac{(-1)^{k+1}(\hbar/2)^{2k}}{m^{2k+1}(2k+1)!} \pi_{1,2} \left[\frac{\partial^{2k+1}U}{\partial x_\alpha^{2k+1}} \frac{1}{f^{1,2}} \frac{\partial^{2k} f^{1,2}}{\partial v_\alpha^{2k}}\right]. \quad (3.27)$$

For a quantum harmonic oscillator with potential $U = \frac{m\omega^2 x^2}{2}$, Wigner function $f^{1,2}$ is known explicitly

$$f^{1,2}_n(x,v) = \frac{(-1)^n m}{\pi\hbar} e^{-\frac{m}{\hbar\omega}(v^2+\omega^2 x^2)} L_n\left(\frac{2m}{\hbar\omega}(v^2+\omega^2 x^2)\right), \quad (3.28)$$

where $n$ is the number of quantum state, and $L_n -$ are Laguerre polynomials. Let us take [16] as function $f^{1,2,3}$

$$f^{1,2,3}_n(x,v,\dot{v}) = \frac{(-1)^n}{2\pi\sigma_x\sigma_v} e^{-\frac{\dot{v}^2}{2\sigma_{\dot{v}}^2} - \frac{v^2}{2\sigma_v^2}} L_n\left(2\left(\frac{\dot{v}^2}{2\sigma_{\dot{v}}^2} + \frac{v^2}{2\sigma_v^2}\right)\right)\delta(\dot{v} + \omega^2 x), \quad (3.29)$$

where $\omega = \dfrac{\sigma_v}{\sigma_x} = \dfrac{\sigma_{\dot{v}}}{\sigma_v}$, $\sigma_x \sigma_v = \dfrac{\hbar}{2m}$. From expression (3.29) it follows that function $f_n^{1,2,3}(x,v,\dot{v})$ is even in $v$. Consequently, according to Theorem 5

$$\langle \ddot{v} \rangle_1 = \pi_1 \langle \dot{v} \rangle_1 = -\omega^2 \pi_1 x = -\omega^2 \langle v \rangle_1, \tag{3.30}$$

where the Vlasov-Moyal approximation (i.9) for a harmonic oscillator is taken into consideration

$$\langle \dot{v} \rangle_{1,2} = \langle \dot{v} \rangle_1 = -\omega^2 x. \tag{3.31}$$

From the second equation in (3.20), it follows that

$$\dfrac{1}{f^{1,2}} \dfrac{\partial P_{11}^3}{\partial v} = \langle \ddot{v} \rangle_{1,2} + \omega^2 \pi_{1,2} x = \langle \ddot{v} \rangle_{1,2} + \omega^2 v. \tag{3.32}$$

Using expression (3.29) we calculate $P_{11}^3(x,v)$ and obtain

$$\begin{aligned}
P_{11}^3(x,v) &= \int_{-\infty}^{+\infty} \left( \dot{v} - \langle \dot{v} \rangle_{1,2} \right)^2 f^{1,2,3} d\dot{v} = \\
&= \dfrac{(-1)^n}{2\pi \sigma_x \sigma_v} \left( \omega^2 x + \langle \dot{v} \rangle_{1,2} \right)^2 e^{-\frac{\omega^2 x^2}{2\sigma_v^2} - \frac{v^2}{2\sigma_v^2}} L_n \left( 2 \left( \dfrac{\omega^2 x^2}{2\sigma_v^2} + \dfrac{v^2}{2\sigma_v^2} \right) \right) = 0,
\end{aligned} \tag{3.33}$$

where approximation (3.31) is taken into consideration. From expressions (3.33) and (3.32), it follows that

$$\langle \ddot{v} \rangle_{1,2} = -\omega^2 v. \tag{3.34}$$

Note that when averaging expression (3.34) over the space of velocities one obtains expression (3.30), which becomes zero due to the parity of Wigner function (3.28).

§4 $H_n$ – **Boltzmann functions**

By analogy with work [6], we construct the Boltzmann $H_n$-functions for the dispersion chain of the Vlasov equations (2.18) - (2.20).

**Definition 5.** *Let function $S^{n_1,...,n_R}$ (2.16) satisfy dispersion chain (2.18) - (2.20), then we define the Boltzmann $H^{n_1,...,n_R}$ -function as*

$$H^{n_1,...,n_R}(t) \stackrel{\text{det}}{=} \dfrac{1}{f^0} \int_{\Omega^{n_1}} ... \int_{\Omega^{n_R}} f^{n_1,...,n_R}\left( \vec{\xi}^{n_1,...,n_R}, t \right) S^{n_1,...,n_R} \prod_{s=1}^{R} d^3 \xi^{k_s} = \left\langle S^{n_1,...,n_R} \right\rangle_0 (t), \tag{4.1}$$

*where functions $S^{n_1,...,n_R}$ and $f^{n_1,...,n_R}$ are related by relation (2.16), $f^0 = N(t)$.*

In a particular case, for distribution functions from the first group: $f^1$, $f^{1,2}$, $f^{1,2,3}$,... the corresponding the Boltzmann $H^{n_1,...,n_R}$ functions go into the Boltzmann $H_1$, $H_2$, $H_3$,...-functions [6]. Function $H_2$ coincides with the known Boltzmann $H$ – function proportional to entropy. In fact, the Boltzmann $H^{n_1,...,n_R}$-functions are an extension of the concept of entropy from the classical phase space $\{\vec{r}, m\vec{v}\}$ to the generalized phase space containing an infinite set of kinematical values of all orders $\{\vec{r}, \vec{v}, \dot{\vec{v}}, \ddot{\vec{v}},...\}$.

**Theorem 6** *Let the distribution functions of the form $f^{n_1,...,n_R}$ be positive and satisfy Vlasov dispersion chain (2.18) - (2.20), then the corresponding Boltzmann $H^{n_1,...,n_R}$ – functions (4.1) satisfy the equations:*

$$1^{st}\ rank$$
$$\pi_0\left[f^0 H^n\right] = -f^0 \left\langle Q_n^n \right\rangle_0. \qquad (4.2)$$

$$2^{nd}\ rank$$
$$\pi_0\left[f^0 H^{n,n+1}\right] = -f^0 \left\langle Q_{n,n+1}^{n+1} \right\rangle_0, \qquad (4.3)$$
$$\pi_0\left[f^0 H^{n,n+k}\right] = -f^0 \left[\left\langle Q_{n,n+k}^n \right\rangle_0 + \left\langle Q_{n,n+k}^{n+k} \right\rangle_0\right].$$

$$3^{rd}\ rank$$
$$\pi_0\left[f^0 H^{n,n+1,n+2}\right] = -f^0 \left\langle Q_{n,n+1,n+2}^{n+2} \right\rangle_0, \qquad (4.4)$$
$$\pi_0\left[f^0 H^{n,n+1,n+1+k}\right] = -f^0 \left(\left\langle Q_{n,n+1,n+1+k}^{n+1} \right\rangle_0 + \left\langle Q_{n,n+1,n+1+k}^{n+1+k} \right\rangle_0\right),$$
$$\pi_0\left[f^0 H^{n,n+s,n+s+1}\right] = -f^0 \left(\left\langle Q_{n,n+s,n+s+1}^n \right\rangle_0 + \left\langle Q_{n,n+s,n+s+1}^{n+s+1} \right\rangle_0\right),$$
$$\pi_0\left[f^0 H^{n,n+s,n+s+k}\right] = -f^0 \left(\left\langle Q_{n,n+s,n+s+k}^n \right\rangle_0 + \left\langle Q_{n,n+s,n+s+k}^{n+s} \right\rangle_0 + \left\langle Q_{n,n+s,n+s+k}^{n+s+k} \right\rangle_0\right).$$

The proof of Theorem 6 is given in Appendix B.

Equations (4.2)-(4.4) are similar in structure with dispersion chain equations (2.18)-(2.20). For constant number of particles $f^0 = N(t) = const$, the time variation of the Boltzmann $H^{n_1,...,n_R}$ – functions is completely determined by mean sources of dissipations $\langle Q \rangle$. If the right-hand side of equations (4.2) - (4.4) is positive, then the function of the Boltzmann $H^{n_1,...,n_R}$ – functions decreases, and if negative, then the Boltzmann $H^{n_1,...,n_R}$ – function increases. In the absence of dissipation sources, the Boltzmann $H^{n_1,...,n_R}$ – function is stationary.

*Thus, one may regard equations (4.2) - (4.4) as an analogue of the Boltzmann $H$ – theorem for the dispersion chain of Vlasov equations in a generalized phase space.* Sources of dissipations $\langle Q \rangle$ are the sources of the production of «generalized entropy», the sign of which is determined by vector field $\left\langle \vec{\xi}^{p+1} \right\rangle_{n_1,...,n_R}$ of higher kinematical values. According to definition (2.16) $Q^p_{n_1,...,n_R} \stackrel{det}{=} \mathrm{div}_{\xi^p} \left\langle \vec{\xi}^{p+1} \right\rangle_{n_1,...,n_R}$, therefore, the source of vector field $\left\langle \vec{\xi}^{p+1} \right\rangle_{n_1,...,n_R}$ will give $Q^p_{n_1,...,n_R} > 0$, and the sink of the vector field $\left\langle \vec{\xi}^{p+1} \right\rangle_{n_1,...,n_R}$ will give $Q^p_{n_1,...,n_R} < 0$. Since the averaging

of $\left\langle Q^p_{n_1,..,n_R} \right\rangle_0$ uses positive probability density function $f^{n_1,...,n_R}$, the sign of $\left\langle Q^p_{n_1,...,n_R} \right\rangle_0$ is completely determined by the sign of sources of dissipations $Q^p_{n_1,..,n_R}$.

**Remark**

Let us make an important comment. The Vlasov equations chain contains kinematic equations that lack information on the dynamics. The sign of dissipation sources $Q^p_{n_1,..,n_R}$ cannot be defined from the Vlasov equations. Defining the sign of values $Q^p_{n_1,...,n_R}$ requires one to cut-off the chain and make a dynamic approximation of mean kinematical value $\left\langle \vec{\xi}^{p+1} \right\rangle_{n_1,..,n_R}$ of the form (i.1)/(i.2), (i.6)/(i.9), (i.11)/(i.12). A similar situation arises in the Boltzmann equation during the dynamic approximation of the collision integral.

Note that in the process of constructing the Vlasov equations chain [1], nowhere was the condition of positive distribution functions imposed. Theorem 6 is formulated for positive distribution functions. Of course, from the standpoint of classical physics, the probability density is a positive function and the requirement of Theorem 6 is natural. When describing quantum systems in a phase space, the Wigner function is used, which determines the quasi-probability of the probabilities. "Quasi" term stands for the presence of negative values in the Wigner function. From the standpoint of quantum mechanics in a phase space, it is required to extend the conditions of Theorem 6 to the case $f^{n_1,...,n_R} \in \mathbb{R}$. Note that the Wigner function satisfies the Moyal equation [17], which coincides with the equation for function $f^{1,2}$ at the chain cut-off with dynamic Vlasov-Moyal approximation (i.9) for kinematical value $\left\langle \dot{\vec{v}} \right\rangle_{1,2}$.

Without restricting the generality, we consider an extension of Theorem 6 for distribution functions $f^{n,n+1} \in \mathbb{R}$ of the second rank.

**Theorem 7.** *Let distribution function $f^{n,n+1} \in \mathbb{R}$ of the second rank satisfy the dispersion chain of the Vlasov equations and may be represented in the form:*

$$f^{n,n+1} = \begin{cases} f^{n,n+1}, \ (n,n+1) \in \Omega_n^+ \times \Omega_{n+1}^+, \\ -\left|f^{n,n+1}\right|, \ (n,n+1) \in \Omega_n^- \times \Omega_{n+1}^-, \\ 0, \ (n,n+1) \in \Sigma_{n,n+1}, \end{cases} \quad (4.5)$$

*where $\Omega_n^+ \times \Omega_{n+1}^+$ and $\Omega_n^- \times \Omega_{n+1}^-$ are regions, in which $f^{n,n+1}$ takes on positive and negative values, respectively. Surface $\Sigma_{n,n+1}$ corresponds to the boundary between domains $\Omega_n^+ \times \Omega_{n+1}^+$ and $\Omega_n^- \times \Omega_{n+1}^-$.*

*Then equation (4.3) $\pi_0\left[f^0 H^{n,n+1}\right] = -f^0 \left\langle Q^{n+1}_{n,n+1} \right\rangle_0$ for function $f^{n,n+1}$ must be replaced by the system of equations:*

$$\begin{cases} \pi_0\left[f^0 \left\langle \ln\left|f^{n,n+1}\right| \right\rangle_0\right] = -f^0 \left\langle Q^{n+1}_{n,n+1} \right\rangle_0, \\ f^0_- = const, \end{cases} \quad (4.6)$$

where $f_-^0$ corresponds to the number of particles or to the probability for the system to be located within domain $\Omega_n^- \times \Omega_{n+1}^-$.

The proof of Theorem 6 is given in Appendix B.

Let's consider some properties of boundary $\Sigma_{n,n+1}$ between domains $\Omega_n^+ \times \Omega_{n+1}^+$ and $\Omega_n^- \times \Omega_{n+1}^-$. Since function $f^{n,n+1}$ depends generally on time, the boundary $\Sigma_{n,n+1}$ also can change its shape. When domain $\Omega_n^- \times \Omega_{n+1}^-$ is not single or not simply connected then there are more than one boundaries $\Sigma_{n,n+1}$. As a physical system evolves, its domains $\Omega_n^- \times \Omega_{n+1}^-$ may fuse into a larger single one with a new common boundary or, alternatively, one domain may be split into several ones with their own boundaries.

Theorem 7 has an important conclusion: probability of presence $f_-^0$ (or the number of particles) in domain $\Omega_n^- \times \Omega_{n+1}^-$ is constant. Over time, the value of $f_-^0$ remains unchanged, i.e., there is no flow of probabilities between domains $\Omega_n^+ \times \Omega_{n+1}^+$ and $\Omega_n^- \times \Omega_{n+1}^-$.

As an example of $f^{n,n+1}$, let us consider Wigner function $f_s^{1,2} = W_s$, where $s$ is the number of the quantum state. According to Hudson's theorem [23], only the Gaussian distribution of the Wigner function is positive. It follows from Theorem 7 that such a state will remain unchanged over time. The states of a quantum system with negative Wigner functions will have the same phase areas of negative domains in time. The simplest example of such a system is the Wigner function of a harmonic oscillator (3.28).

In the particular case for $f^{n,n+1} \geq 0$, system of equations (4.6) transforms into equation (4.3) from Theorem 6.

**Conclusions**

The approach set forth in this work provides new insights into physical processes from the standpoint of the generalized phase space of higher kinematical values. On the one hand, the fundamental nature of the Vlasov chain of equations makes it possible to consider various domains of physics in a unified manner: classical mechanics, statistical physics, continuum mechanics and quantum mechanics.

Each of these domains is a special case of the chain and is determined by what is meant by the distribution function and mean kinematical values. For instance, in continuum mechanics, the distribution function can be understood as the density of matter, in quantum mechanics – as the probability density, in field theory – as the charge density (the probability of detecting a charge) or even the magnetic permittivity function [18].

On the other hand, the fundamental nature of the chain is manifested in the level of realistic description of a physical system. The cut-off of the chain at a certain equation determines the degree of information on the system. The consideration of motion equations not higher than the second order is exhausted when considering electromagnetic radiation (Lorentz equations (i.12)). Quantum mechanics in a phase space uses the apparatus of the Wigner function, which satisfies the second Vlasov equation (Moyal's equation [17]) and contains higher-order dissipations.

Clearly, this work is theoretical, but the resulting equations (3.2), (3.3), (3.8), (3.9), (3.13) - (3.15), (3.17) - (3.19) may be directly used to construct conservative finite-difference algorithms for the computational solution of a number of practical problems of hydrodynamics, continuum mechanics and plasma physics with allowance for electromagnetic radiation [19-22, 14, 15].

**Acknowledgements**

This work was supported by the RFBR No. 18-29-10014. This research has been supported by the Interdisciplinary Scientific and Educational School of Moscow University «Photonic and Quantum Technologies. Digital Medicine».

**Appendix A**

*Proof of Theorem 1*

Let us multiply equation (3.1) by $\vec{\xi}^{n+1}$ and integrate it over phase subspace $\Omega_{n+1}$, and we obtain:

$$\frac{\partial}{\partial t}\int_{\Omega_{n+1}} \xi_\alpha^{n+1} f^{n,n+1} d^3\xi^{n+1} + \frac{\partial}{\partial \xi_\beta^n}\int_{\Omega_{n+1}} \xi_\alpha^{n+1}\xi_\beta^{n+1} f^{n,n+1} d^3\xi^{n+1} + \int_{\Omega_{n+1}} \xi_\alpha^{n+1} \frac{\partial}{\partial \xi_\beta^{n+1}}\left[ f^{n,n+1}\left\langle \xi_\beta^{n+2}\right\rangle_{n,n+1}\right] d^3\xi^{n+1} = 0,$$
(A.1)

where the componentwise form of notation $\vec{\xi}^n = \{\xi_1^n, \xi_3^n, \xi_3^n\}$ is used. The second integral in equation (A.1) may be expressed through expression $P_{\alpha\beta}^{n+1}$ (3.4), we obtain

$$P_{\alpha\beta}^{n+1}(n) \overset{\text{det}}{=} \int_{(\infty)} \left(\xi_\alpha^{n+1} - \left\langle\xi_\alpha^{n+1}\right\rangle_n\right)\left(\xi_\beta^{n+1} - \left\langle\xi_\beta^{n+1}\right\rangle_n\right) f^{n,n+1} d^3\xi^{n+1} = \int_{(\infty)} \xi_\alpha^{n+1}\xi_\beta^{n+1} f^{n,n+1} d^3\xi^{n+1} +$$

$$-\left\langle\xi_\beta^{n+1}\right\rangle_n \int_{(\infty)} \xi_\alpha^{n+1} f^{n,n+1} d^3\xi^{n+1} - \left\langle\xi_\alpha^{n+1}\right\rangle_n \int_{(\infty)} \xi_\beta^{n+1} f^{n,n+1} d^3\xi^{n+1} + \left\langle\xi_\alpha^{n+1}\right\rangle_n \left\langle\xi_\beta^{n+1}\right\rangle_n \int_{(\infty)} f^{n,n+1} d^3\xi^{n+1},$$

$$\int_{\Omega_{n+1}} \xi_\alpha^{n+1}\xi_\beta^{n+1} f^{n,n+1} d^3\xi^{n+1} = P_{\alpha\beta}^{n+1} + f^n \left\langle\xi_\beta^{n+1}\right\rangle_n \left\langle\xi_\alpha^{n+1}\right\rangle_n.$$
(A.2)

where $P_{\alpha\beta}^{n+1}\left(\vec{\xi}^n\right) \overset{\text{det}}{=} P_{\alpha\beta}^{n+1}(n)$. Substituting (A.2) into (A.1), integrating by parts and using formula (1.11), we obtain:

$$\frac{\partial}{\partial t}\left[f^n \left\langle\xi_\alpha^{n+1}\right\rangle_n\right] + \frac{\partial P_{\alpha\beta}^{n+1}}{\partial \xi_\beta^n} + \frac{\partial}{\partial \xi_\beta^n}\left[f^n \left\langle\xi_\beta^{n+1}\right\rangle_n \left\langle\xi_\alpha^{n+1}\right\rangle_n\right] - \int_{\Omega_{n+1}} f^{n,n+1}\left\langle\xi_\alpha^{n+2}\right\rangle_{n,n+1} d^3\xi^{n+1} = 0,$$
(A.3)

where the integration took into consideration the fast decrease of the distribution functions at infinity [1] and the condition $\frac{\partial \xi_\alpha^{n+1}}{\partial \xi_\beta^{n+1}} = \delta_{\alpha\beta}$. Since function $f^n$ of the first rank satisfies equation $\frac{\partial f^n}{\partial t} = -\frac{\partial}{\partial \xi_\beta^n}\left[f^n \left\langle\xi_\beta^{n+1}\right\rangle_n\right]$ (2.13), expression (A.3) takes the form:

$$\pi_n \left\langle\xi_\alpha^{n+1}\right\rangle_n = \left[\frac{\partial}{\partial t} + \left\langle\xi_\beta^{n+1}\right\rangle_n \frac{\partial}{\partial \xi_\beta^n}\right]\left\langle\xi_\alpha^{n+1}\right\rangle_n = -\frac{1}{f^n}\frac{\partial P_{\alpha\beta}^{n+1}}{\partial \xi_\beta^n} + \left\langle\xi_\alpha^{n+2}\right\rangle_n.$$
(A.4)

Let us multiply initial equation (3.1) by $\frac{1}{2}\left(\xi^{n+1}\right)^2$ and integrate over space $\Omega_{n+1}$.

$$\frac{1}{2}\frac{\partial}{\partial t}\int\limits_{(\infty)}\left(\xi^{n+1}\right)^2 f^{n,n+1}d^3\xi^{n+1}+\frac{1}{2}\frac{\partial}{\partial \xi_\beta^n}\int\limits_{(\infty)}\left(\xi^{n+1}\right)^2\xi_\beta^n f^{n,n+1}d^3\xi^{n+1}+$$
$$+\frac{1}{2}\int\limits_{(\infty)}\left(\xi^{n+1}\right)^2\frac{\partial}{\partial \xi_\beta^{n+1}}\left[f^{n,n+1}\left\langle\xi_\beta^{n+2}\right\rangle_{n,n+1}\right]d^3\xi^{n+1}=0, \tag{A.5}$$

Integrating it by parts and considering that $\frac{1}{2}\frac{\partial}{\partial \xi_\beta^{n+1}}\left(\xi_\alpha^{n+1}\xi_\alpha^{n+1}\right)=\xi_\alpha^{n+1}\delta_{\alpha\beta}$, expression (A.5) takes the form

$$\frac{\partial}{\partial t}\left[\frac{f^n}{2}\left\langle\left(\xi^{n+1}\right)^2\right\rangle_n\right]+\frac{\partial}{\partial \xi_\beta^n}\int\limits_{\Omega_{n+1}}\frac{1}{2}\left(\xi^{n+1}\right)^2\xi_\beta^n f^{n,n+1}d^3\xi^{n+1}-\int\limits_{\Omega_{n+1}}f^{n,n+1}\left\langle\xi_\alpha^{n+2}\right\rangle_{n,n+1}\xi_\alpha^{n+1}d^3\xi^{n+1}=0, \tag{A.6}$$

where summation is meant over repeated indices. Let us express the second integral in equation (A.6) in terms of tensor $P_{\alpha\beta\mu}^{n+1}$ (3.5).

$$P_{\alpha\beta\mu}^{n+1}(n)\overset{\text{det}}{=}\int\limits_{\Omega_{n+1}}\left(\xi_\alpha^{n+1}-\left\langle\xi_\alpha^{n+1}\right\rangle_n\right)\left(\xi_\beta^{n+1}-\left\langle\xi_\beta^{n+1}\right\rangle_n\right)\left(\xi_\mu^{n+1}-\left\langle\xi_\mu^{n+1}\right\rangle_n\right)f^{n,n+1}d^3\xi^{n+1}=$$
$$=\int\limits_{\Omega_{n+1}}\xi_\alpha^{n+1}\xi_\beta^{n+1}\xi_\mu^{n+1}f^{n,n+1}d^3\xi^{n+1}-\left\langle\xi_\beta^{n+1}\right\rangle_n\int\limits_{\Omega_{n+1}}\xi_\alpha^{n+1}\xi_\mu^{n+1}f^{n,n+1}d^3\xi^{n+1}-\left\langle\xi_\alpha^{n+1}\right\rangle_n\int\limits_{\Omega_{n+1}}\xi_\beta^{n+1}\xi_\mu^{n+1}f^{n,n+1}d^3\xi^{n+1}+$$
$$+\left\langle\xi_\alpha^{n+1}\right\rangle_n\left\langle\xi_\beta^{n+1}\right\rangle_n\int\limits_{\Omega_{n+1}}\xi_\mu^{n+1}f^{n,n+1}d^3\xi^{n+1}-\left\langle\xi_\mu^{n+1}\right\rangle_n\int\limits_{\Omega_{n+1}}\xi_\alpha^{n+1}\xi_\beta^{n+1}f^{n,n+1}d^3\xi^{n+1}+$$
$$+\left\langle\xi_\beta^{n+1}\right\rangle_n\left\langle\xi_\mu^{n+1}\right\rangle_n\int\limits_{\Omega_{n+1}}\xi_\alpha^{n+1}f^{n,n+1}d^3\xi^{n+1}+\left\langle\xi_\alpha^{n+1}\right\rangle_n\left\langle\xi_\mu^{n+1}\right\rangle_n\int\limits_{\Omega_{n+1}}\xi_\beta^{n+1}f^{n,n+1}d^3\xi^{n+1}-\left\langle\xi_\alpha^{n+1}\right\rangle_n\left\langle\xi_\beta^{n+1}\right\rangle_n\left\langle\xi_\mu^{n+1}\right\rangle_n=$$
$$=\int\limits_{\Omega_{n+1}}\xi_\alpha^{n+1}\xi_\beta^{n+1}\xi_\mu^{n+1}f^{n,n+1}d^3\xi^{n+1}-\left\langle\xi_\beta^{n+1}\right\rangle_n\int\limits_{\Omega_{n+1}}\xi_\alpha^{n+1}\xi_\mu^{n+1}f^{n,n+1}d^3\xi^{n+1}-\left\langle\xi_\alpha^{n+1}\right\rangle_n\int\limits_{\Omega_{n+1}}\xi_\beta^{n+1}\xi_\mu^{n+1}f^{n,n+1}d^3\xi^{n+1}-$$
$$-\left\langle\xi_\mu^{n+1}\right\rangle_n\int\limits_{\Omega_{n+1}}\xi_\alpha^{n+1}\xi_\beta^{n+1}f^{n,n+1}d^3\xi^{n+1}+2f^n\left\langle\xi_\alpha^{n+1}\right\rangle_n\left\langle\xi_\beta^{n+1}\right\rangle_n\left\langle\xi_\mu^{n+1}\right\rangle_n.$$

Considering expression (A.2), we obtain

$$P_{\alpha\beta\mu}^{n+1}=\int\limits_{\Omega_{n+1}}\xi_\alpha^{n+1}\xi_\beta^{n+1}\xi_\mu^{n+1}f^{n,n+1}d^3\xi^{n+1}-\left\langle\xi_\beta^{n+1}\right\rangle_n P_{\alpha\mu}^{n+1}-f^n\left\langle\xi_\mu^{n+1}\right\rangle_n\left\langle\xi_\alpha^{n+1}\right\rangle_n\left\langle\xi_\beta^{n+1}\right\rangle_n-\left\langle\xi_\alpha^{n+1}\right\rangle_n P_{\beta\mu}^{n+1}-$$
$$-f^n\left\langle\xi_\alpha^{n+1}\right\rangle_n\left\langle\xi_\beta^{n+1}\right\rangle_n\left\langle\xi_\mu^{n+1}\right\rangle_n-\left\langle\xi_\mu^{n+1}\right\rangle_n P_{\alpha\beta}^{n+1}-f^n\left\langle\xi_\beta^{n+1}\right\rangle_n\left\langle\xi_\alpha^{n+1}\right\rangle_n\left\langle\xi_\mu^{n+1}\right\rangle_n+2f^n\left\langle\xi_\alpha^{n+1}\right\rangle_n\left\langle\xi_\beta^{n+1}\right\rangle_n\left\langle\xi_\mu^{n+1}\right\rangle_n=$$
$$=\int\limits_{\Omega_{n+1}}\xi_\alpha^{n+1}\xi_\beta^{n+1}\xi_\mu^{n+1}f^{n,n+1}d^3\xi^{n+1}-\left\langle\xi_\beta^{n+1}\right\rangle_n P_{\alpha\mu}^{n+1}-\left\langle\xi_\alpha^{n+1}\right\rangle_n P_{\beta\mu}^{n+1}-\left\langle\xi_\mu^{n+1}\right\rangle_n P_{\alpha\beta}^{n+1}-f^n\left\langle\xi_\mu^{n+1}\right\rangle_n\left\langle\xi_\alpha^{n+1}\right\rangle_n\left\langle\xi_\beta^{n+1}\right\rangle_n,$$

$$\int\limits_{\Omega_{n+1}}\xi_\alpha^{n+1}\xi_\beta^{n+1}\xi_\mu^{n+1}f^{n,n+1}d^3\xi^{n+1}=$$
$$=P_{\alpha\beta\mu}^{n+1}+\left\langle\xi_\beta^{n+1}\right\rangle_n P_{\alpha\mu}^{n+1}+\left\langle\xi_\alpha^{n+1}\right\rangle_n P_{\beta\mu}^{n+1}+\left\langle\xi_\mu^{n+1}\right\rangle_n P_{\alpha\beta}^{n+1}+f^n\left\langle\xi_\mu^{n+1}\right\rangle_n\left\langle\xi_\alpha^{n+1}\right\rangle_n\left\langle\xi_\beta^{n+1}\right\rangle_n. \tag{A.7}$$

Consequently,

$$\int_{\Omega_{n+1}} \left(\xi^{n+1}\right)^2 \xi_\mu^{n+1} f^{n,n+1} d^3\xi^{n+1} = \operatorname{Tr} P_{\alpha\alpha\mu}^{n+1} + 2\left\langle \xi_\alpha^{n+1} \right\rangle_n P_{\alpha\mu}^{n+1} + \left\langle \xi_\mu^{n+1} \right\rangle_n \operatorname{Tr} P_{\alpha\alpha}^{n+1} + f^n \left\langle \xi_\mu^{n+1} \right\rangle_n \left\langle \xi^{n+1} \right\rangle_n^2. \qquad (A.8)$$

Taking into account expression (A.2), the first summand in equation (A.6) takes the form:

$$f^n \left\langle \left(\xi^{n+1}\right)^2 \right\rangle_n = \int_{\Omega_{n+1}} \left(\xi^{n+1}\right)^2 f^{n,n+1} d^3\xi^{n+1} = \operatorname{Tr} P_{\alpha\alpha}^{n+1} + f^n \left\langle \xi^{n+1} \right\rangle_n^2. \qquad (A.9)$$

Substituting (A.9) and (A.8) into equation (A.6), we obtain

$$\frac{\partial}{\partial t}\left[\frac{f^n}{2}\left\langle \xi^{n+1} \right\rangle_n^2 + \frac{1}{2}\operatorname{Tr} P_{\alpha\alpha}^{n+1}\right] + \frac{\partial}{\partial \xi_\beta^n}\left[\frac{f^n}{2}\left\langle \xi^{n+1} \right\rangle_n^2 \left\langle \xi_\beta^{n+1} \right\rangle_n + \frac{1}{2}\left\langle \xi_\beta^{n+1} \right\rangle_n \operatorname{Tr} P_{\alpha\alpha}^{n+1} + \left\langle \xi_\alpha^{n+1} \right\rangle_n P_{\alpha\beta}^{n+1} + \frac{1}{2}\operatorname{Tr} P_{\alpha\alpha\beta}^{n+1}\right] =$$
$$= \int_{\Omega_{n+1}} f^{n,n+1} \left\langle \xi_\alpha^{n+2} \right\rangle_{n,n+1} \xi_\alpha^{n+1} d^3\xi^{n+1}. \qquad (A.10)$$

Theorem 1 is proved.

*Proof of Theorem 2*

Let us multiply equation (3.7) by $\vec{\xi}^{n+k}$ and integrate it over phase subspace $\Omega_{n+k}$, we get:

$$\frac{\partial}{\partial t}\int_{\Omega_{n+k}} \xi_\alpha^{n+k} f^{n,n+k} d^3\xi^{n+k} + \frac{\partial}{\partial \xi_\beta^n}\int_{\Omega_{n+k}} \xi_\alpha^{n+k} \left\langle \xi_\beta^{n+1} \right\rangle_{n,n+k} f^{n,n+k} d^3\xi^{n+k} +$$
$$+ \int_{\Omega_{n+k}} \xi_\alpha^{n+k} \frac{\partial}{\partial \xi_\beta^{n+k}}\left[f^{n,n+k} \left\langle \xi_\beta^{n+k+1} \right\rangle_{n,n+k}\right] d^3\xi^{n+k} = 0,$$

$$\left\langle \xi_\alpha^{n+k} \right\rangle_n \frac{\partial f^n}{\partial t} + f^n \frac{\partial \left\langle \xi_\alpha^{n+k} \right\rangle_n}{\partial t} + \frac{\partial}{\partial \xi_\beta^n}\int_{\Omega_{n+k}} \xi_\alpha^{n+k} \left\langle \xi_\beta^{n+1} \right\rangle_{n,n+k} f^{n,n+k} d^3\xi^{n+k} - \int_{\Omega_{n+k}} f^{n,n+k} \left\langle \xi_\alpha^{n+k+1} \right\rangle_{n,n+k} d^3\xi^{n+k} = 0,$$
$$\qquad (A.11)$$

Using integration by parts in the fourth summand in (A.11) and using equation (2.13), we obtain

$$f^n \frac{\partial \left\langle \xi_\alpha^{n+k} \right\rangle_n}{\partial t} - \left\langle \xi_\alpha^{n+k} \right\rangle_n \frac{\partial}{\partial \xi_\beta^n}\left[f^n \left\langle \xi_\beta^{n+1} \right\rangle_n\right] + \frac{\partial}{\partial \xi_\beta^n}\int_{\Omega_{n+k}} \xi_\alpha^{n+k} \left\langle \xi_\beta^{n+1} \right\rangle_{n,n+k} f^{n,n+k} d^3\xi^{n+k} = f^n \left\langle \xi_\alpha^{n+k+1} \right\rangle_n. \qquad (A.12)$$

Let us express the third summand in equation (A.12) through the momentum of the second order $P_{\alpha\beta}^{n+1,n+k}$ (3.10), determined by kinematical values of different orders.

$$P_{\alpha\beta}^{n+1,n+k}(n) = \int_{\Omega_{n+1}}\int_{\Omega_{n+k}} \left(\xi_\alpha^{n+1} - \left\langle \xi_\alpha^{n+1} \right\rangle_n\right)\left(\xi_\beta^{n+k} - \left\langle \xi_\beta^{n+k} \right\rangle_n\right) f^{n,n+1,n+k} d^3\xi^{n+1} d^3\xi^{n+k} =$$
$$= \int_{\Omega_{n+1}}\int_{\Omega_{n+k}} \xi_\alpha^{n+1} \xi_\beta^{n+k} f^{n,n+1,n+k} d^3\xi^{n+1} d^3\xi^{n+k} - \left\langle \xi_\alpha^{n+1} \right\rangle_n \int_{\Omega_{n+k}} \xi_\beta^{n+k} f^{n,n+k} d^3\xi^{n+k} -$$

$$-\left\langle\xi_\beta^{n+k}\right\rangle_n \int_{\Omega_{n+1}} \xi_\alpha^{n+1} f^{n,n+1} d^3\xi^{n+1} + f^n \left\langle\xi_\alpha^{n+1}\right\rangle_n \left\langle\xi_\beta^{n+k}\right\rangle_n = \int_{\Omega_{n+1}}\int_{\Omega_{n+k}} \xi_\alpha^{n+1}\xi_\beta^{n+k} f^{n,n+1,n+k} d^3\xi^{n+1} d^3\xi^{n+k} - f^n \left\langle\xi_\alpha^{n+1}\right\rangle_n \left\langle\xi_\beta^{n+k}\right\rangle_n ,$$

$$\int_{\Omega_{n+1}}\int_{\Omega_{n+k}} \xi_\alpha^{n+1}\xi_\beta^{n+k} f^{n,n+1,n+k} d^3\xi^{n+1} d^3\xi^{n+k} = P_{\alpha\beta}^{n+1,n+k} + f^n \left\langle\xi_\alpha^{n+1}\right\rangle_n \left\langle\xi_\beta^{n+k}\right\rangle_n . \tag{A.13}$$

from here

$$\int_{\Omega_{n+k}} \xi_\alpha^{n+k} \left\langle\xi_\beta^{n+1}\right\rangle_{n,n+k} f^{n,n+k} d^3\xi^{n+k} = \int_{\Omega_{n+1}}\int_{\Omega_{n+k}} \xi_\alpha^{n+k}\xi_\beta^{n+1} f^{n,n+1,n+k} d^3\xi^{n+1} d^3\xi^{n+k} = $$
$$= P_{\beta\alpha}^{n+1,n+k} + f^n \left\langle\xi_\alpha^{n+k}\right\rangle_n \left\langle\xi_\beta^{n+1}\right\rangle_n . \tag{A.14}$$

Substituting (A.14) into equation (A.12), we obtain

$$f^n \frac{\partial \left\langle\xi_\alpha^{n+k}\right\rangle_n}{\partial t} - \left\langle\xi_\alpha^{n+k}\right\rangle_n \frac{\partial}{\partial \xi_\beta^n}\left[f^n \left\langle\xi_\beta^{n+1}\right\rangle_n\right] + \frac{\partial P_{\beta\alpha}^{n+1,n+k}}{\partial \xi_\beta^n} + \frac{\partial}{\partial \xi_\beta^n}\left[f^n \left\langle\xi_\alpha^{n+k}\right\rangle_n \left\langle\xi_\beta^{n+1}\right\rangle_n\right] = f^n \left\langle\xi_\alpha^{n+k+1}\right\rangle_n ,$$

$$\pi_n \left\langle\xi_\alpha^{n+k}\right\rangle_n = \left[\frac{\partial}{\partial t} + \left\langle\xi_\beta^{n+1}\right\rangle_n \frac{\partial}{\partial \xi_\beta^n}\right]\left\langle\xi_\alpha^{n+k}\right\rangle_n = -\frac{1}{f^n}\frac{\partial P_{\beta\alpha}^{n+1,n+k}}{\partial \xi_\beta^n} + \left\langle\xi_\alpha^{n+k+1}\right\rangle_n . \tag{A.15}$$

Let us construct the equation of the law of conservation of «energy» for the second group of equations. We multiply initial equation (3.7) by $\frac{1}{2}\left(\xi^{n+k}\right)^2$ and integrate over space $\Omega_{n+k}$

$$\frac{\partial}{\partial t}\int_{\Omega_{n+k}} \frac{f^{n,n+k}}{2}\left(\xi^{n+k}\right)^2 d^3\xi^{n+k} + \frac{\partial}{\partial \xi_\beta^n}\int_{\Omega_{n+k}} \left(\xi^{n+k}\right)^2 \frac{f^{n,n+k}}{2} \left\langle\xi_\beta^{n+1}\right\rangle_{n,n+k} d^3\xi^{n+k} + $$
$$+\int_{\Omega_{n+k}} \frac{\left(\xi^{n+k}\right)^2}{2} \frac{\partial}{\partial \xi_\beta^{n+k}}\left[f^{n,n+k}\left\langle\xi_\beta^{n+k+1}\right\rangle_{n,n+k}\right] d^3\xi^{n+k} = 0.$$

$$\frac{\partial}{\partial t}\left[\frac{f^n}{2}\left\langle\left(\xi^{n+k}\right)^2\right\rangle_n\right] + \frac{\partial}{\partial \xi_\beta^n}\int_{\Omega_{n+k}} \left(\xi^{n+k}\right)^2 \frac{f^{n,n+k}}{2} \left\langle\xi_\beta^{n+1}\right\rangle_{n,n+k} d^3\xi^{n+k} = \int_{\Omega_{n+k}} f^{n,n+k}\left\langle\xi_\alpha^{n+k+1}\right\rangle_{n,n+k} \xi_\alpha^{n+k} d^3\xi^{n+k}.$$
(A.16)

Let us transform the second integral in equation (A.16) using momentum of the third order $P_{\beta\alpha\alpha}^{n+1,n+k,n+k}$ (3.11) of mixed kinematical values.

$$P_{\beta\alpha\alpha}^{n+1,n+k,n+k}(n) \stackrel{\text{det}}{=} \int_{\Omega_{n+k}}\int_{\Omega_{n+1}} \left(\xi_\alpha^{n+k} - \left\langle\xi_\alpha^{n+k}\right\rangle_n\right)^2 \left(\xi_\beta^{n+1} - \left\langle\xi_\beta^{n+1}\right\rangle_n\right) f^{n,n+1,n+k} d^3\xi^{n+1} d^3\xi^{n+k} = $$
$$= \int_{\Omega_{n+k}}\int_{\Omega_{n+1}} \xi_\alpha^{n+k}\xi_\alpha^{n+k}\xi_\beta^{n+1} f^{n,n+1,n+k} d^3\xi^{n+1} d^3\xi^{n+k} - 2\left\langle\xi_\alpha^{n+k}\right\rangle_n \int_{\Omega_{n+k}}\int_{\Omega_{n+1}} \xi_\alpha^{n+k}\xi_\beta^{n+1} f^{n,n+1,n+k} d^3\xi^{n+1} d^3\xi^{n+k} + $$
$$+\left\langle\xi_\alpha^{n+k}\right\rangle_n^2 \int_{\Omega_{n+1}} \xi_\beta^{n+1} f^{n,n+1} d^3\xi^{n+1} - \left\langle\xi_\beta^{n+1}\right\rangle_n \int_{\Omega_{n+k}}\int_{\Omega_{n+1}} \xi_\alpha^{n+k}\xi_\alpha^{n+k} f^{n,n+1,n+k} d^3\xi^{n+1} d^3\xi^{n+k} $$
$$+2\left\langle\xi_\beta^{n+1}\right\rangle_n \left\langle\xi_\alpha^{n+k}\right\rangle_n \int_{\Omega_{n+k}} \xi_\alpha^{n+k} f^{n,n+k} d^3\xi^{n+k} - f^n \left\langle\xi_\beta^{n+1}\right\rangle_n \left\langle\xi_\alpha^{n+k}\right\rangle_n^2 .$$

From expressions (A.2) and (A.13), it follows that

$$\int_{\Omega_{n+k}} \int_{\Omega_{n+1}} \xi_\alpha^{n+k} \xi_\alpha^{n+k} f^{n,n+1,n+k} d^3\xi^{n+1} d^3\xi^{n+k} = P_{\alpha\alpha}^{n+k,} + f^n \left\langle \xi_\alpha^{n+k} \right\rangle_n^2,$$

$$\int_{\Omega_{n+k}} \int_{\Omega_{n+1}} \xi_\beta^{n+1} \xi_\alpha^{n+k} f^{n,n+1,n+k} d^3\xi^{n+1} d^3\xi^{n+k} = P_{\beta\alpha}^{n+1,n+k} + f^n \left\langle \xi_\alpha^{n+k} \right\rangle_n \left\langle \xi_\beta^{n+1} \right\rangle_n,$$

from here

$$P_{\beta\alpha\alpha}^{n+1,n+k,n+k} = \int_{\Omega_{n+k}} \int_{\Omega_{n+1}} \xi_\alpha^{n+k} \xi_\alpha^{n+k} \xi_\beta^{n+1} f^{n,n+1,n+k} d^3\xi^{n+1} d^3\xi^{n+k} - 2\left\langle \xi_\alpha^{n+k} \right\rangle_n P_{\beta\alpha}^{n+1,n+k} -$$

$$-2f^n \left\langle \xi_\alpha^{n+k} \right\rangle_n^2 \left\langle \xi_\beta^{n+1} \right\rangle_n - \left\langle \xi_\beta^{n+1} \right\rangle_n P_{\alpha\alpha}^{n+k,} - f^n \left\langle \xi_\beta^{n+1} \right\rangle_n \left\langle \xi_\alpha^{n+k} \right\rangle_n^2 + 2f^n \left\langle \xi_\beta^{n+1} \right\rangle_n \left\langle \xi_\alpha^{n+k} \right\rangle_n^2 =$$

$$= \int_{\Omega_{n+k}} \int_{\Omega_{n+1}} \xi_\alpha^{n+k} \xi_\alpha^{n+k} \xi_\beta^{n+1} f^{n,n+1,n+k} d^3\xi^{n+1} d^3\xi^{n+k} - 2\left\langle \xi_\alpha^{n+k} \right\rangle_n P_{\beta\alpha}^{n+1,n+k} - \left\langle \xi_\beta^{n+1} \right\rangle_n P_{\alpha\alpha}^{n+k} - f^n \left\langle \xi_\beta^{n+1} \right\rangle_n \left\langle \xi_\alpha^{n+k} \right\rangle_n^2,$$

$$\int_{\Omega_{n+k}} \int_{\Omega_{n+1}} \xi_\alpha^{n+k} \xi_\alpha^{n+k} \xi_\beta^{n+1} f^{n,n+1,n+k} d^3\xi^{n+1} d^3\xi^{n+k} =$$

$$= P_{\beta\alpha\alpha}^{n+1,n+k,n+k} + 2\left\langle \xi_\alpha^{n+k} \right\rangle_n P_{\beta\alpha}^{n+1,n+k} + \left\langle \xi_\beta^{n+1} \right\rangle_n P_{\alpha\alpha}^{n+k} + f^n \left\langle \xi_\beta^{n+1} \right\rangle_n \left\langle \xi_\alpha^{n+k} \right\rangle_n^2.$$

(A.17)

Using (A.17), the second summand in equation (A.16) takes the form:

$$\int_{\Omega_{n+k}} \left(\xi^{n+k}\right)^2 f^{n,n+k} \left\langle \xi_\beta^{n+1} \right\rangle_{n,n+k} d^3\xi^{n+k} = \int_{\Omega_{n+1}} \int_{\Omega_{n+k}} \left(\xi^{n+k}\right)^2 \xi_\beta^{n+1} f^{n,n+1,n+k} d^3\xi^{n+k} d^3\xi^{n+1} =$$

$$= \operatorname{Tr} P_{\beta\alpha\alpha}^{n+1,n+k,n+k} + \left\langle \xi_\beta^{n+1} \right\rangle_n \operatorname{Tr} P_{\alpha\alpha}^{n+k,} + 2 P_{\beta\alpha}^{n+1,n+k} \left\langle \xi_\alpha^{n+k} \right\rangle_n + f^n \left\langle \xi_\beta^{n+1} \right\rangle_n \left\langle \xi^{n+k} \right\rangle_n^2.$$

(A.18)

The first summand in equation (A.16) may be expressed in terms of the momentum of the second order (A.2)

$$f^n \left\langle \left(\xi^{n+k}\right)^2 \right\rangle_n = \sum_\alpha \int_{\Omega_{n+k}} \xi_\alpha^{n+k} \xi_\alpha^{n+k} f^{n,n+k} d^3\xi^{n+k} = \operatorname{Tr} P_{\alpha\alpha}^{n+k} + f^n \left\langle \xi^{n+k} \right\rangle_n^2.$$

(A.19)

Substituting (A.18), (A.19) into equation (A.16), we obtain

$$\frac{\partial}{\partial t}\left[\frac{f^n}{2} \left\langle \xi^{n+k} \right\rangle_n^2 + \frac{1}{2}\operatorname{Tr} P_{\alpha\alpha}^{n+k}\right] +$$

$$+ \frac{\partial}{\partial \xi_\beta^n}\left[\frac{f^n}{2}\left\langle \xi^{n+k} \right\rangle_n^2 \left\langle \xi_\beta^{n+1} \right\rangle_n + \frac{1}{2}\left\langle \xi_\beta^{n+1} \right\rangle_n \operatorname{Tr} P_{\alpha\alpha}^{n+k} + P_{\beta\alpha}^{n+1,n+k} \left\langle \xi_\alpha^{n+k} \right\rangle_n + \frac{1}{2}\operatorname{Tr} P_{\beta\alpha\alpha}^{n+1,n+k,n+k}\right] =$$

$$= \int_{\Omega_{n+k}} f^{n,n+k} \left\langle \xi_\alpha^{n+k+1} \right\rangle_{n,n+k} \xi_\alpha^{n+k} d^3\xi^{n+k}.$$

(A.20)

Theorem 2 is proved.

*Proof of Theorem 3*

Let us multiply equation (3.12) by $\vec{\xi}^{n+2}$ and integrate it over space $\Omega_{n+2}$:

$$\frac{\partial}{\partial t}\left[f^{n,n+1}\left\langle\xi_\alpha^{n+2}\right\rangle_{n,n+1}\right]+\frac{\partial}{\partial\xi_\beta^n}\int_{\Omega_{n+2}}f^{n,n+1,n+2}\xi_\alpha^{n+2}\xi_\beta^{n+1}d^3\xi^{n+2}+$$
$$+\frac{\partial}{\partial\xi_\beta^{n+1}}\int_{\Omega_{n+2}}f^{n,n+1,n+2}\xi_\alpha^{n+2}\xi_\beta^{n+2}d^3\xi^{n+2}=f^{n,n+1}\left\langle\xi_\alpha^{n+3}\right\rangle_{n,n+1}.$$
(A.21)

The first integral in equation (A.21) equals $\left\langle\xi_\alpha^{n+2}\right\rangle_{n,n+1}\xi_\beta^{n+1}f^{n,n+1}$. We express the value of the second integral in terms of the momentum of the second order for kinematical values of the same order. If distribution function $f^{n,n+1,n+2}$ is of the third rank then, by analogy with (A.2), tensor $P_{\alpha\beta}^{n+2}(n,n+1)$ for kinematical values of order $n+2$ takes the form:

$$P_{\alpha\beta}^{n+2}(n,n+1)\stackrel{\text{det}}{=}\int_{\Omega_{n+2}}\left(\xi_\alpha^{n+2}-\left\langle\xi_\alpha^{n+2}\right\rangle_{n,n+1}\right)\left(\xi_\beta^{n+2}-\left\langle\xi_\beta^{n+2}\right\rangle_{n,n+1}\right)f^{n,n+1,n+2}d^3\xi^{n+2}=$$
$$=\int_{\Omega_{n+2}}\xi_\alpha^{n+2}\xi_\beta^{n+2}f^{n,n+1,n+2}d^3\xi^{n+2}-f^{n,n+1}\left\langle\xi_\beta^{n+2}\right\rangle_{n,n+1}\left\langle\xi_\alpha^{n+2}\right\rangle_{n,n+1},$$
(A.22)

where $P_{\alpha\beta}^{n+2}\left(\vec{\xi}^n,\vec{\xi}^{n+1}\right)\stackrel{\text{det}}{=}P_{\alpha\beta}^{n+2}(n,n+1)$. We express partial derivative $\frac{\partial}{\partial t}f^{n,n+1}$ from equation (3.1)

$$\frac{\partial f^{n,n+1}}{\partial t}=-\xi_\beta^{n+1}\frac{\partial f^{n,n+1}}{\partial\xi_\beta^n}-\frac{\partial}{\partial\xi_\beta^{n+1}}\left[f^{n,n+1}\left\langle\xi_\beta^{n+2}\right\rangle_{n,n+1}\right].$$
(A.23)

Substituting (A.22) and (A.23) into equation (A.21), we obtain:

$$\pi_{n,n+1}\left\langle\xi_\alpha^{n+2}\right\rangle_{n,n+1}=\left[\frac{\partial}{\partial t}+\xi_\beta^{n+1}\frac{\partial}{\partial\xi_\beta^n}+\left\langle\xi_\beta^{n+2}\right\rangle_{n,n+1}\frac{\partial}{\partial\xi_\beta^{n+1}}\right]\left\langle\xi_\alpha^{n+2}\right\rangle_{n,n+1}=-\frac{1}{f^{n,n+1}}\frac{\partial P_{\alpha\beta}^{n+2}}{\partial\xi_\beta^{n+1}}+\left\langle\xi_\alpha^{n+3}\right\rangle_{n,n+1}.$$
(A.24)

Let us multiply equation (3.12) by value $\vec{\xi}^n$ and integrate it over $\Omega_n$, we obtain

$$\frac{\partial}{\partial t}\left[f^{n+1,n+2}\left\langle\xi_\alpha^n\right\rangle_{n+1,n+2}\right]+\frac{\partial}{\partial\xi_\beta^{n+1}}\left[\left\langle\xi_\alpha^n\right\rangle_{n+1,n+2}\xi_\beta^{n+2}f^{n+1,n+2}\right]+$$
$$+\frac{\partial}{\partial\xi_\beta^{n+2}}\int_{\Omega_n}\xi_\alpha^n\left\langle\xi_\beta^{n+3}\right\rangle_{n,n+1,n+2}f^{n,n+1,n+2}d^3\xi^n=\xi_\alpha^{n+1}f^{n+1,n+2}.$$
(A.25)

Considering that

$$\int_{\Omega_n}\xi_\alpha^n\left\langle\xi_\beta^{n+3}\right\rangle_{n,n+1,n+2}f^{n,n+1,n+2}d^3\xi^n=\int_{\Omega_{n+3}}\int_{\Omega_n}\xi_\alpha^n\xi_\beta^{n+3}f^{n,n+1,n+2,n+3}d^3\xi^nd^3\xi^{n+3}=$$
$$=P_{\alpha\beta}^{n,n+3}(n+1,n+2)+f^{n+1,n+2}\left\langle\xi_\alpha^n\right\rangle_{n+1,n+2}\left\langle\xi_\beta^{n+3}\right\rangle_{n+1,n+2},$$
(A.26)

$$\frac{\partial f^{n+1,n+2}}{\partial t}=-\xi_\beta^{n+2}\frac{\partial f^{n+1,n+2}}{\partial\xi_\beta^{n+1}}-\frac{\partial}{\partial\xi_\beta^{n+2}}\left[f^{n+1,n+2}\left\langle\xi_\beta^{n+3}\right\rangle_{n+1,n+2}\right],$$

equation (A.25) takes the form:

$$f^{n+1,n+2}\frac{\partial \langle \xi_\alpha^n \rangle_{n+1,n+2}}{\partial t} + f^{n+1,n+2}\xi_\beta^{n+2}\frac{\partial}{\partial \xi_\beta^{n+1}}\langle \xi_\alpha^n \rangle_{n+1,n+2} + f^{n+1,n+2}\langle \xi_\beta^{n+3}\rangle_{n+1,n+2}\frac{\partial}{\partial \xi_\beta^{n+2}}\langle \xi_\alpha^n \rangle_{n+1,n+2} =$$

$$= \xi_\alpha^{n+1}f^{n+1,n+2} - \frac{\partial P_{\alpha\beta}^{n,n+3}}{\partial \xi_\beta^{n+2}},$$

$$\left[\frac{\partial}{\partial t} + \xi_\beta^{n+2}\frac{\partial}{\partial \xi_\beta^{n+1}} + \langle \xi_\beta^{n+3}\rangle_{n+1,n+2}\frac{\partial}{\partial \xi_\beta^{n+2}}\right]\langle \xi_\alpha^n \rangle_{n+1,n+2} = -\frac{1}{f^{n+1,n+2}}\frac{\partial P_{\alpha\beta}^{n,n+3}}{\partial \xi_\beta^{n+2}} + \xi_\alpha^{n+1}. \qquad (A.27)$$

Let us multiply equation (3.12) by value $\vec{\xi}^{n+1}$ and integrate in over $\Omega_{n+1}$, we obtain

$$\frac{\partial}{\partial t}\left[\langle \xi_\alpha^{n+1}\rangle_{n,n+2}f^{n,n+2}\right] + \frac{\partial}{\partial \xi_\beta^n}\int_{\Omega_{n+1}}\xi_\alpha^{n+1}\xi_\beta^{n+1}f^{n,n+1,n+2}d^3\xi^{n+1} +$$

$$+ \frac{\partial}{\partial \xi_\beta^{n+2}}\int_{\Omega_{n+1}}\xi_\alpha^{n+1}\langle \xi_\beta^{n+3}\rangle_{n,n+1,n+2}f^{n,n+1,n+2}d^3\xi^{n+1} = \qquad (A.28)$$

$$= \xi_\alpha^{n+2}f^{n,n+2}.$$

Taking into account

$$\int_{\Omega_{n+1}}\xi_\alpha^{n+1}\xi_\beta^{n+1}f^{n,n+1,n+2}d^3\xi^{n+1} = P_{\alpha\beta}^{n+1}(n,n+2) + f^{n,n+2}\langle \xi_\alpha^{n+1}\rangle_{n,n+2}\langle \xi_\beta^{n+1}\rangle_{n,n+2}, \qquad (A.29)$$

$$\int_{\Omega_{n+1}}\xi_\alpha^{n+1}\langle \xi_\beta^{n+3}\rangle_{n,n+1,n+2}f^{n,n+1,n+2}d^3\xi^{n+1} = \int_{\Omega_{n+3}}\int_{\Omega_{n+1}}\xi_\alpha^{n+1}\xi_\beta^{n+3}f^{n,n+1,n+2,n+3}d^3\xi^{n+1}d^3\xi^{n+3},$$

$$\int_{\Omega_{n+3}}\int_{\Omega_{n+1}}\xi_\alpha^{n+1}\xi_\beta^{n+3}f^{n,n+1,n+2,n+3}d^3\xi^{n+1}d^3\xi^{n+3} = P_{\alpha\beta}^{n+1,n+3}(n,n+2) + f^{n,n+2}\langle \xi_\alpha^{n+1}\rangle_{n,n+2}\langle \xi_\beta^{n+3}\rangle_{n,n+2},$$

$$\frac{\partial f^{n,n+2}}{\partial t} = -\frac{\partial}{\partial \xi_\beta^n}\left[\langle \xi_\beta^{n+1}\rangle_{n,n+2}f^{n,n+2}\right] - \frac{\partial}{\partial \xi_\beta^{n+2}}\left[f^{n,n+2}\langle \xi_\beta^{n+3}\rangle_{n,n+2}\right],$$

we obtain

$$f^{n,n+2}\frac{\partial}{\partial t}\langle \xi_\alpha^{n+1}\rangle_{n,n+2} + f^{n,n+2}\langle \xi_\beta^{n+1}\rangle_{n,n+2}\frac{\partial}{\partial \xi_\beta^n}\langle \xi_\alpha^{n+1}\rangle_{n,n+2} + f^{n,n+2}\langle \xi_\beta^{n+3}\rangle_{n,n+2}\frac{\partial}{\partial \xi_\beta^{n+2}}\langle \xi_\alpha^{n+1}\rangle_{n,n+2} =$$

$$= -\frac{\partial P_{\alpha\beta}^{n+1}}{\partial \xi_\beta^n} - \frac{\partial P_{\alpha\beta}^{n+1,n+3}}{\partial \xi_\beta^{n+2}} + \xi_\alpha^{n+2}f^{n,n+2},$$

$$\left[\frac{\partial}{\partial t} + \langle \xi_\beta^{n+1}\rangle_{n,n+2}\frac{\partial}{\partial \xi_\beta^n} + \langle \xi_\beta^{n+3}\rangle_{n,n+2}\frac{\partial}{\partial \xi_\beta^{n+2}}\right]\langle \xi_\alpha^{n+1}\rangle_{n,n+2} = -\frac{1}{f^{n,n+2}}\left(\frac{\partial P_{\alpha\beta}^{n+1}}{\partial \xi_\beta^n} + \frac{\partial P_{\alpha\beta}^{n+1,n+3}}{\partial \xi_\beta^{n+2}}\right) + \xi_\alpha^{n+2}. \qquad (A.30)$$

Theorem 3 is proved.

*Proof of Theorem 4*

Multiplying expression (3.16) by $\vec{\xi}^{n+1+k}$ and integrating over subspace $\Omega_{n+1+k}$, we obtain:

$$\frac{\partial}{\partial t}\left[\left\langle \xi_\alpha^{n+1+k}\right\rangle_{n,n+1} f^{n,n+1}\right] + \xi_\beta^{n+1}\frac{\partial}{\partial \xi_\beta^n}\left[\left\langle \xi_\alpha^{n+1+k}\right\rangle_{n,n+1} f^{n,n+1}\right] +$$
$$+\frac{\partial}{\partial \xi_\beta^{n+1}}\int_{\Omega_{n+1+k}} f^{n,n+1,n+1+k}\xi_\alpha^{n+1+k}\left\langle \xi_\beta^{n+2}\right\rangle_{n,n+1,n+1+k} d^3\xi^{n+1+k} = f^{n,n+1}\left\langle \xi_\alpha^{n+2+k}\right\rangle_{n,n+1}.$$
(A.31)

Considering that

$$\int_{\Omega_{n+1+k}} f^{n,n+1,n+1+k}\xi_\alpha^{n+1+k}\left\langle \xi_\beta^{n+2}\right\rangle_{n,n+1,n+1+k} d^3\xi^{n+1+k} = \int_{\Omega_{n+2}}\int_{\Omega_{n+1+k}} f^{n,n+1,n+2,n+1+k}\xi_\alpha^{n+1+k}\xi_\beta^{n+2} d^3\xi^{n+1+k} d^3\xi^{n+2}$$
$$= P_{\alpha\beta}^{n+2,n+1+k}(n,n+1) + f^{n,n+1}\left\langle \xi_\beta^{n+k+1}\right\rangle_{n,n+1}\left\langle \xi_\alpha^{n+2}\right\rangle_{n,n+1},$$

equation (A.31) takes the form:

$$f^{n,n+1}\frac{\partial\left\langle \xi_\alpha^{n+1+k}\right\rangle_{n,n+1}}{\partial t} + \left\langle \xi_\alpha^{n+1+k}\right\rangle_{n,n+1}\frac{\partial f^{n,n+1}}{\partial t} + \xi_\beta^{n+1}\frac{\partial}{\partial \xi_\beta^n}\left[\left\langle \xi_\alpha^{n+1+k}\right\rangle_{n,n+1} f^{n,n+1}\right] +$$
$$+\frac{\partial}{\partial \xi_\beta^{n+1}}\left[f^{n,n+1}\left\langle \xi_\alpha^{n+k+1}\right\rangle_{n,n+1}\left\langle \xi_\beta^{n+2}\right\rangle_{n,n+1}\right] = -\frac{\partial P_{\beta\alpha}^{n+2,n+1+k}}{\partial \xi_\beta^{n+1}} + f^{n,n+1}\left\langle \xi_\alpha^{n+2+k}\right\rangle_{n,n+1},$$
$$\left[\frac{\partial}{\partial t} + \xi_\beta^{n+1}\frac{\partial}{\partial \xi_\beta^n} + \left\langle \xi_\beta^{n+2}\right\rangle_{n,n+1}\frac{\partial}{\partial \xi_\beta^{n+1}}\right]\left\langle \xi_\alpha^{n+k+1}\right\rangle_{n,n+1} = -\frac{1}{f^{n,n+1}}\frac{\partial P_{\beta\alpha}^{n+2,n+1+k}}{\partial \xi_\beta^{n+1}} + \left\langle \xi_\alpha^{n+2+k}\right\rangle_{n,n+1}. \quad (A.32)$$

Multiplying equation (3.16) by $\vec{\xi}^{n+1}$ and integrating over subspace $\Omega_{n+1}$, we obtain:

$$\frac{\partial}{\partial t}\left[\left\langle \xi_\alpha^{n+1}\right\rangle_{n,n+1+k} f^{n,n+1+k}\right] + \frac{\partial}{\partial \xi_\beta^n}\int_{\Omega_{n+1}}\xi_\alpha^{n+1}\xi_\beta^{n+1} f^{n,n+1,n+1+k} d^3\xi^{n+1} +$$
$$+\frac{\partial}{\partial \xi_\beta^{n+1+k}}\int_{\Omega_{n+1}} f^{n,n+1,n+1+k}\xi_\alpha^{n+1}\left\langle \xi_\beta^{n+2+k}\right\rangle_{n,n+1,n+1+k} d^3\xi^{n+1} = f^{n,n+1+k}\left\langle \xi_\alpha^{n+2}\right\rangle_{n,n+1+k}.$$
(A.33)

Considering that

$$\int_{\Omega_{n+1}} f^{n,n+1,n+1+k}\xi_\alpha^{n+1}\left\langle \xi_\beta^{n+2+k}\right\rangle_{n,n+1,n+1+k} d^3\xi^{n+1} = \int_{\Omega_{n+2+k}}\int_{\Omega_{n+1}} f^{n,n+1,n+1+k,n+2+k}\xi_\alpha^{n+1}\xi_\beta^{n+2+k} d^3\xi^{n+1} d^3\xi^{n+2+k} =$$
$$= P_{\alpha\beta}^{n+1,n+2+k}(n,n+1+k) + f^{n,n+1+k}\left\langle \xi_\alpha^{n+1}\right\rangle_{n,n+1+k}\left\langle \xi_\beta^{n+2+k}\right\rangle_{n,n+1+k}, \quad (A.34)$$

$$\int_{\Omega_{n+1}}\xi_\alpha^{n+1}\xi_\beta^{n+1} f^{n,n+1,n+1+k} d^3\xi^{n+1} = P_{\alpha\beta}^{n+1}(n+1+k) + f^{n,n+1+k}\left\langle \xi_\alpha^{n+1}\right\rangle_{n,n+1+k}\left\langle \xi_\beta^{n+1}\right\rangle_{n,n+1+k},$$

$$\frac{\partial f^{n,n+1+k}}{\partial t} = -\frac{\partial}{\partial \xi_\beta^n}\left[\left\langle \xi_\beta^{n+1}\right\rangle_{n,n+1+k} f^{n,n+1+k}\right] - \frac{\partial}{\partial \xi_\beta^{n+1+k}}\left[f^{n,n+1+k}\left\langle \xi_\beta^{n+2+k}\right\rangle_{n,n+1+k}\right],$$

equation (A.33) takes the form:

$$f^{n,n+1+k}\frac{\partial \langle \xi_\alpha^{n+1}\rangle_{n,n+1+k}}{\partial t} + \langle \xi_\alpha^{n+1}\rangle_{n,n+1+k}\frac{\partial f^{n,n+1+k}}{\partial t} + \frac{\partial P_{\alpha\beta}^{n+1}}{\partial \xi_\beta^n} + \frac{\partial}{\partial \xi_\beta^n}\left[ f^{n,n+1+k}\langle \xi_\alpha^{n+1}\rangle_{n,n+1+k}\langle \xi_\beta^{n+1}\rangle_{n,n+1+k}\right] +$$

$$+ \frac{\partial P_{\alpha\beta}^{n+1,n+2+k}}{\partial \xi_\beta^{n+1+k}} + \frac{\partial}{\partial \xi_\beta^{n+1+k}}\left[ f^{n,n+1+k}\langle \xi_\alpha^{n+1}\rangle_{n,n+1+k}\langle \xi_\beta^{n+2+k}\rangle_{n,n+1+k}\right] = f^{n,n+1+k}\langle \xi_\alpha^{n+2}\rangle_{n,n+1+k}.$$

$$\left[\frac{\partial}{\partial t} + \langle \xi_\beta^{n+1}\rangle_{n,n+1+k}\frac{\partial}{\partial \xi_\beta^n} + \langle \xi_\beta^{n+2+k}\rangle_{n,n+1+k}\frac{\partial}{\partial \xi_\beta^{n+1+k}}\right]\langle \xi_\alpha^{n+1}\rangle_{n,n+1+k} =$$

$$= -\frac{1}{f^{n,n+1+k}}\left(\frac{\partial P_{\alpha\beta}^{n+1}}{\partial \xi_\beta^n} + \frac{\partial P_{\alpha\beta}^{n+1,n+2+k}}{\partial \xi_\beta^{n+1+k}}\right) + \langle \xi_\alpha^{n+2}\rangle_{n,n+1+k}.$$  (A.35)

Multiplying expression (3.16) by $\vec{\xi}^n$ and integrating over subspace $\Omega_n$, we obtain:

$$f^{n+1,n+1+k}\frac{\partial \langle \xi_\alpha^n\rangle_{n+1,n+1+k}}{\partial t} - \langle \xi_\alpha^n\rangle_{n+1,n+1+k}\frac{\partial}{\partial \xi_\beta^{n+1}}\left[ f^{n+1,n+1+k}\langle \xi_\beta^{n+2}\rangle_{n+1,n+1+k}\right] -$$

$$-\langle \xi_\alpha^n\rangle_{n+1,n+1+k}\frac{\partial}{\partial \xi_\beta^{n+1+k}}\left[ f^{n+1,n+1+k}\langle \xi_\beta^{n+2+k}\rangle_{n+1,n+1+k}\right] + \frac{\partial}{\partial \xi_\beta^{n+1}}\left[ f^{n+1,n+1+k}\langle \xi_\alpha^n\rangle_{n+1,n+1+k}\langle \xi_\beta^{n+2}\rangle_{n+1,n+1+k}\right] +$$

$$+\frac{\partial}{\partial \xi_\beta^{n+1+k}}\left[ f^{n+1,n+1+k}\langle \xi_\alpha^n\rangle_{n+1,n+1+k}\langle \xi_\beta^{n+2+k}\rangle_{n+1,n+1+k}\right] = -\frac{\partial P_{\alpha\beta}^{n,n+2}}{\partial \xi_\beta^{n+1}} - \frac{\partial P_{\alpha\beta}^{n,n+2+k}}{\partial \xi_\beta^{n+1+k}} + f^{n+1,n+1+k}\xi_\alpha^{n+1},$$  (A.36)

where the following is taken into account

$$\int_{\Omega_n}\xi_\alpha^n \langle \xi_\beta^{n+2}\rangle_{n,n+1,n+1+k} f^{n,n+1,n+1+k} d^3\xi^n = \int_{\Omega_{n+2}}\int_{\Omega_n}\xi_\alpha^n \xi_\beta^{n+2} f^{n,n+1,n+2,n+1+k} d^3\xi^n d^3\xi^{n+2} =$$

$$= P_{\alpha\beta}^{n,n+2}(n+1,n+1+k) + f^{n+1,n+1+k}\langle \xi_\alpha^n\rangle_{n+1,n+1+k}\langle \xi_\beta^{n+2}\rangle_{n+1,n+1+k},$$

$$\int_{\Omega_n} f^{n,n+1,n+1+k}\langle \xi_\beta^{n+2+k}\rangle_{n,n+1,n+1+k} d^3\xi^n = \int_{\Omega_{n+2+k}}\int_{\Omega_n}\xi_\alpha^n \xi_\beta^{n+2+k} f^{n,n+1,n+1+k,n+2+k} d^3\xi^n d^3\xi^{n+2+k} =$$

$$= P_{\alpha\beta}^{n,n+2+k}(n+1,n+1+k) + f^{n+1,n+1+k}\langle \xi_\alpha^n\rangle_{n+1,n+1+k}\langle \xi_\beta^{n+2+k}\rangle_{n+1,n+1+k},$$

$$\frac{\partial f^{n+1,n+1+k}}{\partial t} = -\frac{\partial}{\partial \xi_\beta^{n+1}}\left[ f^{n+1,n+1+k}\langle \xi_\beta^{n+2}\rangle_{n+1,n+1+k}\right] - \frac{\partial}{\partial \xi_\beta^{n+1+k}}\left[ f^{n+1,n+1+k}\langle \xi_\beta^{n+2+k}\rangle_{n+1,n+1+k}\right].$$

Finally, expression (A.36) takes the form:

$$\left[\frac{\partial}{\partial t} + \langle \xi_\beta^{n+2}\rangle_{n+1,n+1+k}\frac{\partial}{\partial \xi_\beta^{n+1}} + \langle \xi_\beta^{n+2+k}\rangle_{n+1,n+1+k}\frac{\partial}{\partial \xi_\beta^{n+1+k}}\right]\langle \xi_\alpha^n\rangle_{n+1,n+1+k} =$$

$$= -\frac{1}{f^{n+1,n+1+k}}\left(\frac{\partial P_{\alpha\beta}^{n,n+2}}{\partial \xi_\beta^{n+1}} + \frac{\partial P_{\alpha\beta}^{n,n+2+k}}{\partial \xi_\beta^{n+1+k}}\right) + \xi_\alpha^{n+1}.$$

Theorem 4 is proved.

*Proof of Theorem 5.*

By definition (3.4), tensor $P_{\alpha\beta}^{n+1+\lambda}\left(\vec{\xi}^n,...,\vec{\xi}^{n+\lambda}\right)$ has the form:

$$P_{\alpha\beta}^{n+1+\lambda}\left(\vec{\xi}^n,...,\vec{\xi}^{n+\lambda}\right) =$$
$$= \int_{\Omega_{n+1+\lambda}} \left(\xi_\alpha^{n+1+\lambda} - \left\langle \xi_\alpha^{n+1+\lambda} \right\rangle_{n,...,n+\lambda}\right)\left(\xi_\beta^{n+1+\lambda} - \left\langle \xi_\beta^{n+1+\lambda} \right\rangle_{n,...,n+\lambda}\right) f^{n,...,n+1+\lambda} d^3\xi^{n+1+\lambda}, \quad \text{(A.37)}$$

where

$$f^{n,...,n+\lambda}\left\langle \xi_\alpha^{n+1+\lambda} \right\rangle_{n,...,n+\lambda} = \int_{\Omega_{n+1+\lambda}} \xi_\alpha^{n+1+\lambda} f^{n,...,n+1+\lambda} d^3\xi^{n+1+\lambda}. \quad \text{(A.38)}$$

Let us demonstrate that $\left\langle \xi_\alpha^{n+1+\lambda} \right\rangle_{n,...,n+\lambda}$ is an even function with respect to variable $\vec{\xi}^{n+\lambda}$. Indeed, from definition (A.38) and the parity of distribution function $f^{n,...,n+1+\lambda}$ (3.21), it follows that

$$f^{n,...,-(n+\lambda)}\left\langle \xi_\alpha^{n+1+\lambda} \right\rangle_{n,...,-(n+\lambda)} = \int_{\Omega_{n+1+\lambda}} \xi_\alpha^{n+1+\lambda} f^{n,...,-(n+\lambda),n+1+\lambda} d^3\xi^{n+1+\lambda} = \int_{\Omega_{n+1+\lambda}} \xi_\alpha^{n+1+\lambda} f^{n,...,n+\lambda,n+1+\lambda} d^3\xi^{n+1+\lambda},$$

$$\left\langle \xi_\alpha^{n+1+\lambda} \right\rangle_{n,...,-(n+\lambda)} = \left\langle \xi_\alpha^{n+1+\lambda} \right\rangle_{n,...,n+\lambda}, \quad \text{(A.39)}$$

where, for compactness of writing, index «$-(n+\lambda)$» corresponds to argument $-\vec{\xi}^{n+\lambda}$. Calculating expression (A.37), we take into consideration condition (A.39), we obtain

$$P_{\alpha\beta}^{n+1+\lambda}\left(\vec{\xi}^n,...,-\vec{\xi}^{n+\lambda}\right) =$$
$$= \int_{\Omega_{n+1+\lambda}} \left(\xi_\alpha^{n+1+\lambda} - \left\langle \xi_\alpha^{n+1+\lambda} \right\rangle_{n,...,-(n+\lambda)}\right)\left(\xi_\beta^{n+1+\lambda} - \left\langle \xi_\beta^{n+1+\lambda} \right\rangle_{n,...,-(n+\lambda)}\right) f^{n,...-(n+\lambda),n+1+\lambda} d^3\xi^{n+1+\lambda} = \quad \text{(A.40)}$$
$$= \int_{\Omega_{n+1+\lambda}} \left(\xi_\alpha^{n+1+\lambda} - \left\langle \xi_\alpha^{n+1+\lambda} \right\rangle_{n,...,n+\lambda}\right)\left(\xi_\beta^{n+1+\lambda} - \left\langle \xi_\beta^{n+1+\lambda} \right\rangle_{n,...,n+\lambda}\right) f^{n,...n+\lambda,n+1+\lambda} d^3\xi^{n+1+\lambda} = P_{\alpha\beta}^{n+1+\lambda}\left(\vec{\xi}^n,...,\vec{\xi}^{n+\lambda}\right).$$

Consequently, $P_{\alpha\beta}^{n+1+\lambda}$ is an even function with respect to variable $\vec{\xi}^{n+\lambda}$. We integrate equation (3.20) over subspace $\Omega_{n+\lambda}$ and take into account the condition of parity (A.40) and obtain

$$\int_{\Omega_{n+\lambda}} \frac{\partial P_{\alpha\beta}^{n+1+\lambda}}{\partial \xi_\beta^{n+\lambda}}\left(\vec{\xi}^n,...,\vec{\xi}^{n+\lambda}\right) d^3\xi^{n+\lambda} = 0 = f^{n,...,n+\lambda-1}\left[\left\langle \xi_\alpha^{n+2+\lambda} \right\rangle_{n,...,n+\lambda-1} - \pi_{n,...,n+\lambda-1}\left\langle \xi_\alpha^{n+1+\lambda} \right\rangle_{n,...,n+\lambda-1}\right],$$

which was to be proved.

**Appendix B**

*Proof of Theorem 6*

Let us consider distribution functions $f^n$ of the first rank, $S^n = \text{Ln } f^n$, satisfying equation (2.18). We multiply equation (2.18) by $(1+S^n)$ and integrate over subspace $\Omega_n$, we obtain

$$\int_{\Omega_n}(1+\ln f^n)\frac{\partial f^n}{\partial t}d^3\xi^n + \int_{\Omega_n}(1+\ln f^n)\nabla_{\xi^n}\left[f^n\left\langle\vec{\xi}^{n+1}\right\rangle_n\right]d^3\xi^n = 0. \tag{B.1}$$

We transform the first integral in expression (B.1):

$$\int_{\Omega_n}(1+\ln f^n)\frac{\partial f^n}{\partial t}d^3\xi^n = \int_{\Omega_n}\frac{\partial}{\partial t}\left(f^n\ln f^n\right)d^3\xi^n = \frac{\partial}{\partial t}\int_{\Omega_n}f^n S^n d^3\xi^n = \frac{d}{dt}\left[f^0(t)\langle S^n\rangle_0(t)\right]. \tag{B.2}$$

The second integral in equation (B.1) has the form:

$$\int_{\Omega_n}(1+S^n)\frac{\partial}{\partial\xi_\beta^n}\left[f^n\left\langle\xi_\beta^{n+1}\right\rangle_n\right]d^3\xi^n = \int_{\Omega_n}\left[f^n\frac{\partial\left\langle\xi_\beta^{n+1}\right\rangle_n}{\partial\xi_\beta^n} + S^n f^n\frac{\partial\left\langle\xi_\beta^{n+1}\right\rangle_n}{\partial\xi_\beta^n} + \left\langle\xi_\beta^{n+1}\right\rangle_n\frac{\partial(f^n S^n)}{\partial\xi_\beta^n}\right]d^3\xi^n =$$

$$= f^0\langle Q_n^n\rangle_0 + \int_{\Omega_n}\frac{\partial}{\partial\xi_\beta^n}\left[S^n f^n\left\langle\xi_\beta^{n+1}\right\rangle_n\right]d^3\xi^n = f^0\langle Q_n^n\rangle_0, \tag{B.3}$$

Substituting expressions (B.2) and (B.3) into equation (B.1), we obtain

$$\frac{d}{dt}\left[f^0\langle S^n\rangle_0\right] = -f^0\langle Q_n^n\rangle_0. \tag{B.4}$$

Distribution function $f^{n,n+1}$ of the second rank from the first group satisfies equation (3.1). We multiply equation (3.1) by $(1+S^{n,n+1})$ and integrate over subspaces $\Omega_n$ and $\Omega_{n+1}$, we obtain:

$$\int_{\Omega_{n+1}}\int_{\Omega_n}(1+S^{n,n+1})\frac{\partial f^{n,n+1}}{\partial t}d^3\xi^n d^3\xi^{n+1} + \int_{\Omega_{n+1}}\int_{\Omega_n}(1+S^{n,n+1})\frac{\partial}{\partial\xi_\beta^n}\left[f^{n,n+1}\xi_\beta^{n+1}\right]d^3\xi^n d^3\xi^{n+1} +$$

$$+\int_{\Omega_{n+1}}\int_{\Omega_n}(1+S^{n,n+1})\frac{\partial}{\partial\xi_\beta^{n+1}}\left[f^{n,n+1}\left\langle\xi_\beta^{n+2}\right\rangle_{n,n+1}\right]d^3\xi^n d^3\xi^{n+1} = 0. \tag{B.5}$$

The first integral in equation (B.5) has the form:

$$\int_{\Omega_n}\int_{\Omega_{n+1}}(1+S^{n,n+1})\frac{\partial f^{n,n+1}}{\partial t}d^3\xi^n d^3\xi^{n+1} = \frac{d}{dt}\int_{\Omega_n}\int_{\Omega_{n+1}}f^{n,n+1}S^{n,n+1}d^3\xi^n d^3\xi^{n+1} = \frac{d}{dt}\left[f^0\langle S^{n,n+1}\rangle_0\right]. \tag{B.6}$$

We transform the expressions under integral sign in the second and third integrals. Let us take into account the following relations

$$\frac{\partial}{\partial\xi_\beta^n}\left[S^{n,n+1}f^{n,n+1}\xi_\beta^{n+1}\right] = \xi_\beta^{n+1}\left(S^{n,n+1}\frac{\partial f^{n,n+1}}{\partial\xi_\beta^n} + f^{n,n+1}\frac{\partial S^{n,n+1}}{\partial\xi_\beta^n}\right) = (1+S^{n,n+1})\frac{\partial}{\partial\xi_\beta^n}\left[f^{n,n+1}\xi_\beta^{n+1}\right], \tag{B.7}$$

$$\frac{\partial}{\partial \xi_\beta^{n+1}}\left[S^{n,n+1} f^{n,n+1} \left\langle \xi_\beta^{n+2} \right\rangle_{n,n+1}\right] = \left\langle \xi_\beta^{n+2} \right\rangle_{n,n+1} f^{n,n+1} \frac{\partial S^{n,n+1}}{\partial \xi_\beta^{n+1}} + \left\langle \xi_\beta^{n+2} \right\rangle_{n,n+1} S^{n,n+1} \frac{\partial f^{n,n+1}}{\partial \xi_\beta^{n+1}} +$$

$$+ S^{n,n+1} f^{n,n+1} \frac{\partial \left\langle \xi_\beta^{n+2} \right\rangle_{n,n+1}}{\partial \xi_\beta^{n+1}} = \left\langle \xi_\beta^{n+2} \right\rangle_{n,n+1} \frac{\partial f^{n,n+1}}{\partial \xi_\beta^{n+1}} + f^{n,n+1} \frac{\partial \left\langle \xi_\beta^{n+2} \right\rangle_{n,n+1}}{\partial \xi_\beta^{n+1}} - f^{n,n+1} \frac{\partial \left\langle \xi_\beta^{n+2} \right\rangle_{n,n+1}}{\partial \xi_\beta^{n+1}} +$$

$$+ S^{n,n+1} \frac{\partial}{\partial \xi_\beta^{n+1}}\left[\left\langle \xi_\beta^{n+2} \right\rangle_{n,n+1} f^{n,n+1}\right],$$

$$\frac{\partial}{\partial \xi_\beta^{n+1}}\left[S^{n,n+1} f^{n,n+1} \left\langle \xi_\beta^{n+2} \right\rangle_{n,n+1}\right] + f^{n,n+1} Q_{n,n+1}^{n+1} = \left(1 + S^{n,n+1}\right)\frac{\partial}{\partial \xi_\beta^{n+1}}\left[f^{n,n+1} \left\langle \xi_\beta^{n+2} \right\rangle_{n,n+1}\right]. \quad \text{(B.8)}$$

Substituting expressions (B.6)-(B.8) into equation (B.5), we obtain

$$\frac{d}{dt}\left[f^0 \left\langle S^{n,n+1} \right\rangle_0\right] + \int_{\Omega_{n+1}} \int_{\Omega_n} \frac{\partial}{\partial \xi_\beta^n}\left[S^{n,n+1} f^{n,n+1} \xi_\beta^{n+1}\right] d^3\xi^n d^3\xi^{n+1} +$$

$$+ \int_{\Omega_{n+1}} \int_{\Omega_n} \frac{\partial}{\partial \xi_\beta^{n+1}}\left[S^{n,n+1} f^{n,n+1} \left\langle \xi_\beta^{n+2} \right\rangle_{n,n+1}\right] d^3\xi^n d^3\xi^{n+1} + \int_{\Omega_{n+1}} \int_{\Omega_n} f^{n,n+1} Q_{n,n+1}^{n+1} d^3\xi^n d^3\xi^{n+1} = 0,$$

$$\frac{d}{dt}\left[f^0 \left\langle S^{n,n+1} \right\rangle_0\right] = -f^0 \left\langle Q_{n,n+1}^{n+1} \right\rangle_0. \quad \text{(B.9)}$$

Let us consider the distribution functions $f^{n,n+k}$, $k > 1$ of the second rank from the second group satisfying equation (3.7). We multiply equation (3.7) by $\left(1 + S^{n,n+k}\right)$ and integrate over subspaces $\Omega_n$ and $\Omega_{n+k}$, we obtain:

$$\int_{\Omega_{n+k}} \int_{\Omega_n} \left(1 + S^{n,n+k}\right)\frac{\partial f^{n,n+k}}{\partial t} d^3\xi^n d^3\xi^{n+k} + \int_{\Omega_{n+k}} \int_{\Omega_n} \left(1 + S^{n,n+k}\right)\frac{\partial}{\partial \xi_\beta^n}\left[f^{n,n+k} \left\langle \xi_\beta^{n+1} \right\rangle_{n,n+k}\right] d^3\xi^n d^3\xi^{n+k} +$$

$$+ \int_{\Omega_{n+1}} \int_{\Omega_n} \left(1 + S^{n,n+k}\right)\frac{\partial}{\partial \xi_\beta^{n+k}}\left[f^{n,n+k} \left\langle \xi_\beta^{n+k+1} \right\rangle_{n,n+k}\right] d^3\xi^n d^3\xi^{n+k} = 0. \quad \text{(B.10)}$$

The first integral is calculated the same way as in expression (B.6). The expressions standing in the second and third integrals may be represented in the form

$$\left(1 + S^{n,n+k}\right)\frac{\partial}{\partial \xi_\beta^n}\left[f^{n,n+k} \left\langle \xi_\beta^{n+1} \right\rangle_{n,n+k}\right] = \frac{\partial}{\partial \xi_\beta^n}\left[S^{n,n+k} f^{n,n+k} \left\langle \xi_\beta^{n+1} \right\rangle_{n,n+k}\right] + f^{n,n+k} Q_{n,n+k}^n, \quad \text{(B.11)}$$

$$\left(1 + S^{n,n+k}\right)\frac{\partial}{\partial \xi_\beta^{n+k}}\left[f^{n,n+k} \left\langle \xi_\beta^{n+2} \right\rangle_{n,n+k}\right] = \frac{\partial}{\partial \xi_\beta^{n+k}}\left[S^{n,n+k} f^{n,n+k} \left\langle \xi_\beta^{n+2} \right\rangle_{n,n+k}\right] + f^{n,n+k} Q_{n,n+k}^{n+k}.$$

Substituting expressions (B.11) into equation (B.10), we obtain

$$\frac{d}{dt}\left[f^0 \left\langle S^{n,n+k} \right\rangle_0\right] = -f^0 \left[\left\langle Q_{n,n+k}^n \right\rangle_0 + \left\langle Q_{n,n+k}^{n+k} \right\rangle_0\right]. \quad \text{(B.12)}$$

For distribution functions $f^{n,n+1,n+2}$ of the third rank from the first group equation (3.12) is satisfied. We multiply equation (3.12) by $\left(1 + S^{n,n+1,n+2}\right)$ and integrate over subspaces $\Omega_n$, $\Omega_{n+1}$ and $\Omega_{n+2}$

$$\frac{d}{dt}\left[f^0\left\langle S^{n,n+1,n+2}\right\rangle_0\right] + \int_{\Omega_{n+2}}\int_{\Omega_{n+1}}\int_{\Omega_n}\left(1+S^{n,n+1,n+2}\right)\xi_\beta^{n+1}\frac{\partial f^{n,n+1,n+2}}{\partial \xi_\beta^n}d^3\xi^n d^3\xi^{n+1}d^3\xi^{n+2} +$$

$$+\int_{\Omega_{n+2}}\int_{\Omega_{n+1}}\int_{\Omega_n}\left(1+S^{n,n+1,n+2}\right)\xi_\beta^{n+2}\frac{\partial f^{n,n+1,n+2}}{\partial \xi_\beta^{n+1}}d^3\xi^n d^3\xi^{n+1}d^3\xi^{n+2} + \quad (B.13)$$

$$+\int_{\Omega_{n+2}}\int_{\Omega_{n+1}}\int_{\Omega_n}\left(1+S^{n,n+1,n+2}\right)\frac{\partial}{\partial \xi_\beta^{n+2}}\left[f^{n,n+1,n+2}\left\langle \xi_\beta^{n+3}\right\rangle_{n,n+1,n+2}\right]d^3\xi^n d^3\xi^{n+1}d^3\xi^{n+2} = 0.$$

By analogy with expressions (B.7) and (B.8) the following relations may be obtained

$$\left(1+S^{n,n+1,n+2}\right)\xi_\beta^{n+2}\frac{\partial f^{n,n+1,n+2}}{\partial \xi_\beta^{n+1}} = \frac{\partial}{\partial \xi_\beta^{n+1}}\left[S^{n,n+1,n+2}f^{n,n+1,n+2}\xi_\beta^{n+2}\right],$$

$$\left(1+S^{n,n+1,n+2}\right)\xi_\beta^{n+1}\frac{\partial f^{n,n+1,n+2}}{\partial \xi_\beta^n} = \frac{\partial}{\partial \xi_\beta^n}\left[S^{n,n+1,n+2}f^{n,n+1,n+2}\xi_\beta^{n+1}\right], \quad (B.14)$$

$$\left(1+S^{n,n+1,n+2}\right)\frac{\partial}{\partial \xi_\beta^{n+2}}\left[f^{n,n+1,n+2}\left\langle \xi_\beta^{n+3}\right\rangle_{n,n+1,n+2}\right] = \frac{\partial}{\partial \xi_\beta^{n+2}}\left[S^{n,n+1,n+2}f^{n,n+1,n+2}\left\langle \xi_\beta^{n+3}\right\rangle_{n,n+1,n+2}\right] +$$
$$+ f^{n,n+1,n+2}Q^{n+2}_{n,n+1,n+2}$$

With the use of representations (B.14), equation (B.13) takes the form:

$$\frac{d}{dt}\left[f^0\left\langle S^{n,n+1,n+2}\right\rangle_0\right] = -f^0\left\langle Q^{n+2}_{n,n+1,n+2}\right\rangle_0. \quad (B.15)$$

Let us consider functions $f^{n,n+1,n+1+k}$, $f^{n,n+s,n+s+1}$ and $f^{n,n+s,n+s+k}$ of the third rank from the second group. We perform the calculations for functions $f^{n,n+1,n+1+k}$ and $f^{n,n+s,n+s+1}$ similar to (B.10), (B.11).

$$\frac{d}{dt}\left[f^0\left\langle S^{n,n+1,n+1+k}\right\rangle_0\right] + \int_{\Omega_{n+1+k}}\int_{\Omega_{n+1}}\int_{\Omega_n}\left(1+S^{n,n+1,n+1+k}\right)\frac{\partial}{\partial \xi_\beta^n}\left[\xi_\beta^{n+1}f^{n,n+1,n+1+k}\right]d^3\xi^n d^3\xi^{n+1}d^3\xi^{n+1+k} +$$

$$+\int_{\Omega_{n+1+k}}\int_{\Omega_{n+1}}\int_{\Omega_n}\left(1+S^{n,n+1,n+1+k}\right)\frac{\partial}{\partial \xi_\beta^{n+1}}\left[f^{n,n+1,n+1+k}\left\langle \xi_\beta^{n+2}\right\rangle_{n,n+1,n+1+k}\right]d^3\xi^n d^3\xi^{n+1}d^3\xi^{n+1+k} + \quad (B.16)$$

$$+\int_{\Omega_{n+1+k}}\int_{\Omega_{n+1}}\int_{\Omega_n}\left(1+S^{n,n+1,n+1+k}\right)\frac{\partial}{\partial \xi_\beta^{n+1+k}}\left[f^{n,n+1,n+1+k}\left\langle \xi_\beta^{n+2+k}\right\rangle_{n,n+1,n+1+k}\right]d^3\xi^n d^3\xi^{n+1}d^3\xi^{n+1+k} = 0.$$

Let us perform the intermediate transformations:

$$\left(1+S^{n,n+1,n+1+k}\right)\frac{\partial}{\partial \xi_\beta^n}\left[\xi_\beta^{n+1}f^{n,n+1,n+1+k}\right] = \frac{\partial}{\partial \xi_\beta^n}\left[S^{n,n+1,n+1+k}f^{n,n+1,n+1+k}\xi_\beta^{n+1}\right],$$

$$\left(1+S^{n,n+1,n+1+k}\right)\frac{\partial}{\partial \xi_\beta^{n+1}}\left[f^{n,n+1,n+1+k}\left\langle \xi_\beta^{n+2}\right\rangle_{n,n+1,n+1+k}\right] =$$
$$= \frac{\partial}{\partial \xi_\beta^{n+1}}\left[S^{n,n+1,n+1+k}f^{n,n+1,n+1+k}\left\langle \xi_\beta^{n+2}\right\rangle_{n,n+1,n+1+k}\right] + f^{n,n+1,n+1+k}Q^{n+1}_{n,n+1,n+1+k}, \quad (B.17)$$

$$\left(1+S^{n,n+1,n+1+k}\right)\frac{\partial}{\partial \xi_\beta^{n+1+k}}\left[f^{n,n+1,n+1+k}\left\langle \xi_\beta^{n+2+k}\right\rangle_{n,n+1,n+1+k}\right]=$$

$$=\frac{\partial}{\partial \xi_\beta^{n+1+k}}\left[S^{n,n+1,n+1+k} f^{n,n+1,n+1+k}\left\langle \xi_\beta^{n+2+k}\right\rangle_{n,n+1,n+1+k}\right]+f^{n,n+1,n+1+k} Q_{n,n+1,n+1+k}^{n+1+k}.$$

Substituting (B.17) into equation (B.16), we obtain

$$\frac{d}{dt}\left[f^0\left\langle S^{n,n+1,n+1+k}\right\rangle_0\right]=-f^0\left(\left\langle Q_{n,n+1,n+1+k}^{n+1}\right\rangle_0+\left\langle Q_{n,n+1,n+1+k}^{n+1+k}\right\rangle_0\right). \tag{B.18}$$

Function $f^{n,n+s,n+s+1}$ satisfies the equation

$$\frac{\partial f^{n,n+s,n+s+1}}{\partial t}+\frac{\partial}{\partial \xi_\beta^n}\left[f^{n,n+s,n+s+1}\left\langle \xi_\beta^{n+1}\right\rangle_{n,n+s,n+s+1}\right]+\frac{\partial}{\partial \xi_\beta^{n+s}}\left[\xi_\beta^{n+s+1} f^{n,n+s,n+s+1}\right]+$$

$$+\frac{\partial}{\partial \xi_\beta^{n+s+1}}\left[f^{n,n+s,n+s+1}\left\langle \xi_\beta^{n+s+2}\right\rangle_{n,n+s,n+s+1}\right]=0, \tag{B.19}$$

which is similar to the equation for function $f^{n,n+1,n+1+k}$, therefore

$$\frac{d}{dt}\left[f^0\left\langle S^{n,n+s,n+s+1}\right\rangle_0\right]=-f^0\left(\left\langle Q_{n,n+s,n+s+1}^{n}\right\rangle_0+\left\langle Q_{n,n+s,n+s+1}^{n+s+1}\right\rangle_0\right). \tag{B.20}$$

Equation (2.20) for function $f^{n,n+s,n+s+k}$ has three sources of dissipations

$$\pi_{n,n+s,n+s+k} S^{n,n+s,n+s+k}=-\left(Q_{n,n+s,n+s+k}^{n}+Q_{n,n+s,n+s+k}^{n+s}+Q_{n,n+s,n+s+k}^{n+s+k}\right),$$

therefore, performing the calculations similar to (B.16) and (B.17), we obtain

$$\frac{d}{dt}\left[f^0\left\langle S^{n,n+s,n+s+k}\right\rangle_0\right]=-f^0\left(\left\langle Q_{n,n+s,n+s+k}^{n}\right\rangle_0+\left\langle Q_{n,n+s,n+s+k}^{n+s}\right\rangle_0+\left\langle Q_{n,n+s,n+s+k}^{n+s+k}\right\rangle_0\right). \tag{B.21}$$

Theorem 6 is proved.

*Proof of Theorem 7*

From expression (4.6), it follows that function $S^{n,n+1}$ (2.16) has complex values

$$S^{n,n+1}=\operatorname{Ln} f^{n,n+1}=\ln\left|f^{n,n+1}\right|+i\arg f^{n,n+1}=\begin{cases}\ln f^{n,n+1},\ (n,n+1)\in\Omega_n^+\times\Omega_{n+1}^+,\\ \ln\left|f^{n,n+1}\right|+i\pi,\ (n,n+1)\in\Omega_n^-\times\Omega_{n+1}^-,\\ \infty,\ (n,n+1)\in\Sigma_{n,n+1}.\end{cases} \tag{B.22}$$

Performing the calculations similar to as for the derivation of equation (4.3) for function $f^{n,n+1}\in\mathbb{R}$ with consideration of (4.6) and (B.22), we obtain

$$\int\limits_{\Omega_{n+1}}\int\limits_{\Omega_n}\left(1+S^{n,n+1}\right)\frac{\partial f^{n,n+1}}{\partial t}d^3\xi^n d^3\xi^{n+1}+\int\limits_{\Omega_{n+1}}\int\limits_{\Omega_n}\left(1+S^{n,n+1}\right)\frac{\partial}{\partial \xi_\beta^n}\left[f^{n,n+1}\xi_\beta^{n+1}\right]d^3\xi^n d^3\xi^{n+1}+$$

$$+\int\limits_{\Omega_{n+1}}\int\limits_{\Omega_n}\left(1+S^{n,n+1}\right)\frac{\partial}{\partial \xi_\beta^{n+1}}\left[f^{n,n+1}\left\langle\xi_\beta^{n+2}\right\rangle_{n,n+1}\right]d^3\xi^n d^3\xi^{n+1}=0.$$
(B.23)

The first integral in equation (B.23) has the form:

$$\int\limits_{\Omega_n\times\Omega_{n+1}}\left(1+S^{n,n+1}\right)\frac{\partial f^{n,n+1}}{\partial t}d^3\xi^n d^3\xi^{n+1}=\frac{d}{dt}\int\limits_{\Omega_n\times\Omega_{n+1}}f^{n,n+1}S^{n,n+1}d^3\xi^n d^3\xi^{n+1}=$$

$$=\frac{d}{dt}\int\limits_{\Omega_n^+\times\Omega_{n+1}^+}f^{n,n+1}S^{n,n+1}d^3\xi^n d^3\xi^{n+1}+\frac{d}{dt}\int\limits_{\Omega_n^-\times\Omega_{n+1}^-}f^{n,n+1}\left(\ln\left|f^{n,n+1}\right|+i\pi\right)d^3\xi^n d^3\xi^{n+1}=$$

$$=\frac{d}{dt}\int\limits_{\Omega_n^+\times\Omega_{n+1}^+}f^{n,n+1}\ln\left|f^{n,n+1}\right|d^3\xi^n d^3\xi^{n+1}+\frac{d}{dt}\int\limits_{\Omega_n^-\times\Omega_{n+1}^-}f^{n,n+1}\ln\left|f^{n,n+1}\right|d^3\xi^n d^3\xi^{n+1}+$$

$$+i\pi\frac{d}{dt}\int\limits_{\Omega_n^-\times\Omega_{n+1}^-}f^{n,n+1}d^3\xi^n d^3\xi^{n+1}=\frac{d}{dt}\int\limits_{\Omega_n\times\Omega_{n+1}}f^{n,n+1}\ln\left|f^{n,n+1}\right|d^3\xi^n d^3\xi^{n+1}+i\pi\frac{d}{dt}f_-^0=$$

$$\int\limits_{\Omega_n\times\Omega_{n+1}}\left(1+S^{n,n+1}\right)\frac{\partial f^{n,n+1}}{\partial t}d^3\xi^n d^3\xi^{n+1}=\frac{d}{dt}f^0\left\langle\ln\left|f^{n,n+1}\right|\right\rangle_0+i\pi\frac{d}{dt}f_-^0,$$
(B.24)

where $f_-^0$ corresponds to the number of particles or to the probability for the system to be located within domain $\Omega_n^-\times\Omega_{n+1}^-$.

Let us transform the expressions under integral sign in the second and third integrals. We take into account relations (B.7) and (B.8). Integrating each term of expression (B.8) over subspace $\Omega_n\times\Omega_{n+1}$, we obtain

$$\int\limits_{\Omega_n\times\Omega_{n+1}}\frac{\partial}{\partial \xi_\beta^{n+1}}\left[S^{n,n+1}f^{n,n+1}\left\langle\xi_\beta^{n+2}\right\rangle_{n,n+1}\right]d^3\xi^n d^3\xi^{n+1}=\int\limits_{\Omega_n^+\times\Omega_{n+1}^+}\frac{\partial}{\partial \xi_\beta^{n+1}}\left[\left\langle\xi_\beta^{n+2}\right\rangle_{n,n+1}f^{n,n+1}\ln\left|f^{n,n+1}\right|\right]d^3\xi^n d^3\xi^{n+1}+$$

$$+\int\limits_{\Omega_n^-\times\Omega_{n+1}^-}\frac{\partial}{\partial \xi_\beta^{n+1}}\left[\left\langle\xi_\beta^{n+2}\right\rangle_{n,n+1}f^{n,n+1}\ln\left|f^{n,n+1}\right|\right]d^3\xi^n d^3\xi^{n+1}+i\pi\int\limits_{\Omega_n^-\times\Omega_{n+1}^-}\frac{\partial}{\partial \xi_\beta^{n+1}}\left[\left\langle\xi_\beta^{n+2}\right\rangle_{n,n+1}f^{n,n+1}\right]d^3\xi^n d^3\xi^{n+1}=$$

$$=\int\limits_{\Omega_n\times\Omega_{n+1}}\frac{\partial}{\partial \xi_\beta^{n+1}}\left[\left\langle\xi_\beta^{n+2}\right\rangle_{n,n+1}f^{n,n+1}\ln\left|f^{n,n+1}\right|\right]d^3\xi^n d^3\xi^{n+1}+i\pi\int\limits_{\Omega_n^-\times\Omega_{n+1}^-}\frac{\partial}{\partial \xi_\beta^{n+1}}\left[\left\langle\xi_\beta^{n+2}\right\rangle_{n,n+1}f^{n,n+1}\right]d^3\xi^n d^3\xi^{n+1}=$$

$$=i\pi\int\limits_{\Omega_n^-\times\Omega_{n+1}^-}\frac{\partial}{\partial \xi_\beta^{n+1}}\left[\left\langle\xi_\beta^{n+2}\right\rangle_{n,n+1}f^{n,n+1}\right]d^3\xi^n d^3\xi^{n+1},$$

$$\int\limits_{\Omega_n\times\Omega_{n+1}}\frac{\partial}{\partial \xi_\beta^{n+1}}\left[S^{n,n+1}f^{n,n+1}\left\langle\xi_\beta^{n+2}\right\rangle_{n,n+1}\right]d^3\xi^n d^3\xi^{n+1}=i\pi\int\limits_{\Omega_n^-}\int\limits_{\Sigma_{n,n+1}}\xi_\beta^{n+1}f^{n,n+1}d\sigma_\beta^{n+1}d^3\xi^n=0,$$
(B.25)

where, according to (4.6) $\left.f^{n,n+1}\right|_{\Sigma_{n,n+1}}=0$. The second integral of expression (B.8)

$$\int\limits_{\Omega_n\times\Omega_{n+1}}f^{n,n+1}Q_{n,n+1}^{n+1}d^3\xi^n d^3\xi^{n+1}=f^0\left\langle Q_{n,n+1}^{n+1}\right\rangle_0.$$
(B.26)

By analogy with (B.25), the integral of expression (B.7) takes the form:

$$\int_{\Omega_n \times \Omega_{n+1}} \frac{\partial}{\partial \xi_\beta^n} \Big[ S^{n,n+1} f^{n,n+1} \xi_\beta^{n+1} \Big] d^3\xi^n d^3\xi^{n+1} = \int_{\Omega_n^+ \times \Omega_{n+1}^+} \frac{\partial}{\partial \xi_\beta^n} \Big[ \xi_\beta^{n+1} f^{n,n+1} \ln \big| f^{n,n+1} \big| \Big] d^3\xi^n d^3\xi^{n+1} +$$

$$+ \int_{\Omega_n^- \times \Omega_{n+1}^-} \frac{\partial}{\partial \xi_\beta^n} \Big[ \xi_\beta^{n+1} f^{n,n+1} \ln \big| f^{n,n+1} \big| \Big] d^3\xi^n d^3\xi^{n+1} + i\pi \int_{\Omega_n^- \times \Omega_{n+1}^-} \frac{\partial}{\partial \xi_\beta^n} \Big[ \xi_\beta^{n+1} f^{n,n+1} \Big] d^3\xi^n d^3\xi^{n+1} =$$

$$= \int_{\Omega_n \times \Omega_{n+1}} \frac{\partial}{\partial \xi_\beta^n} \Big[ \xi_\beta^{n+1} f^{n,n+1} \ln \big| f^{n,n+1} \big| \Big] d^3\xi^n d^3\xi^{n+1} + i\pi \int_{\Omega_n^- \times \Omega_{n+1}^-} \frac{\partial}{\partial \xi_\beta^n} \Big[ \xi_\beta^{n+1} f^{n,n+1} \Big] d^3\xi^n d^3\xi^{n+1} =$$

$$= i\pi \int_{\Omega_n^- \times \Omega_{n+1}^-} \frac{\partial}{\partial \xi_\beta^n} \Big[ \xi_\beta^{n+1} f^{n,n+1} \Big] d^3\xi^n d^3\xi^{n+1},$$

$$\int_{\Omega_n \times \Omega_{n+1}} \frac{\partial}{\partial \xi_\beta^n} \Big[ S^{n,n+1} f^{n,n+1} \xi_\beta^{n+1} \Big] d^3\xi^n d^3\xi^{n+1} = i\pi \int_{\Omega_{n+1}^-} \int_{\Sigma_{n,n+1}} \xi_\beta^{n+1} f^{n,n+1} d\sigma_\beta^n d^3\xi^{n+1} = 0, \qquad (B.27)$$

where according to (4.6) $f^{n,n+1}\big|_{\Sigma_{n,n+1}} = 0$.

Substituting (B.24) - (B.27) into equation (B.23), we obtain

$$\frac{d}{dt} f^0 \big\langle \ln \big| f^{n,n+1} \big| \big\rangle_0 + i\pi \frac{d}{dt} f_-^0 + f^0 \big\langle Q_{n,n+1}^{n+1} \big\rangle_0 = 0, \qquad (B.28)$$

that is

$$\begin{cases} \dfrac{d}{dt} f^0 \big\langle \ln \big| f^{n,n+1} \big| \big\rangle_0 = -f^0 \big\langle Q_{n,n+1}^{n+1} \big\rangle_0, \\ \dfrac{d}{dt} f_-^0 = 0. \end{cases}$$

Theorem 7 is proved.